%% file: nsnv2.tex
\documentclass[prb,aps,epsfig,twocolumn,showpacs]{revtex4}

\usepackage{epsfig,amssymb,amsmath,latexsym}

\input prbmacro

\begin{document}

\textheight=23.8cm

\title{Renormalization group study of transport through a
superconducting\\
junction of multiple one-dimensional quantum wires}

\author{Sourin \surname{Das}}
\email[Email: ]{sourin.das@weizmann.ac.il}
\affiliation{%
Department of Condensed Matter Physics, \\
Weizmann Institute of Science, Rehovot 76100, Israel
}%
\author{Sumathi \surname{Rao}}
\email[Email: ]{sumathi@hri.res.in}
\author{Arijit \surname{Saha}}
\email[Email: ]{arijit@hri.res.in} \affiliation{Harish-Chandra
Research Institute, \\ Chhatnag Road, Jhusi, Allahabad 211019,
India
}%

\date{\today}
\pacs{73.23.-b,74.45.+c,71.10.Pm}

\begin{abstract}
We investigate transport properties of a superconducting junction of
many ($N \ge 2$) one-dimensional quantum wires. We include the
effect of electron-electron interaction within the one-dimensional
quantum wire using a weak interaction renormalization group
procedure. Due to the proximity effect, transport across the
junction occurs via direct tunneling as well as via the crossed
Andreev channel. We find that the fixed point structure of this
system is far more rich than the fixed point structure of a normal
metal$-$superconductor junction ($N = 1$), where we only have two
fixed points -  the fully insulating fixed point or the Andreev
fixed point. Even a two wire ($N=2$) system  with a superconducting
junction \ie~ a normal metal$-$superconductor$-$normal metal
structure, has non-trivial fixed points with intermediate
transmissions and reflections. We also include electron-electron
interaction induced back-scattering in the quantum wires in our
study and hence obtain non-Luttinger liquid behaviour. It is
interesting to note that  {\textsl{(a)}} effects due to inclusion of
electron-electron interaction induced back-scattering in the wire,
and {\textsl{(b)}} competition between the charge transport via the
electron and hole channels across the junction, give rise to a
non-monotonic behavior of conductance as a function of temperature.
We also find that transport across the junction depends on two
independent interaction parameters. The first one is due to the
usual correlations coming from Friedel oscillations for spin-full
electrons giving rise to the well-known interaction parameter
(${{\alpha = (g_2-2g_1)/2 \pi \hbar v_F}}$).
The second one arises due to the scattering
induced by the proximity  of the superconductor and is given by
(${{\alpha^\prime = (g_2 + g_1)/2 \pi \hbar v_F}}$).
The non-monotonic conductance and the
identification of this new interaction parameter are two of
our main results.
In both the expressions ${{g_1 = V(2k_F)}}$, ${{g_2 = V(0)}}$, where $V(k)$
is the inter electron interaction potential.
\end{abstract}

\maketitle
\section{\label{sec:intro}Introduction}

Effects due to the proximity of a superconductor has motivated a lot
of work~\cite{andreev,blonder,beenaker} in the last several
decades. A direct manifestation of proximity effect is the
 phenomenon of Andreev reflection (\ard) in which
an electron like quasi-particle incident on
{\textsl{normal$-$superconductor}} (\nsd) junction is reflected back
as a hole along with the transfer of two electrons into the
superconductor as a Cooper pair. An even more intriguing example
where the proximity effect manifests itself is the phenomenon of
crossed Andreev reflection (\card) which can only take place in a
{\textsl{normal metal$-$superconductor$-$normal metal}} (\nsnd)
junction provided the distance between the two normal metals is less
than or equal to the phase coherence length of the superconductor.
This is a nonlocal process where an incident electron from one of
the normal leads pairs up with an electron from the other lead to
form a Cooper pair~\cite{byers,feinberg,exp,hekking1,hekking2} and
joins the superconductor. Its relevance in the manipulation of spin
currents~\cite{drsahaepl} (\scd) and questions regarding production
of entangled electron pairs in nano devices for quantum computation
has attracted a lot of attention in recent
times~\cite{buttiker1,buttiker2,buttiker3,recher,bouchiat,beckmann,
russo,chandrasekhar,yeyati}. Further extensions such as inclusion of
effects due to electron-electron interactions on \ar processes in case of
\ns junctions in the context of one-dimensional (\odd) wires have also
been considered recently~\cite{jap1,jap3,fazio,bena,man,titov}.

In this article, we shall first develop a general formulation for
studying the transport properties of a multiple quantum wire
(\qwd) junction in the spirit of the
{\textquotedblleft}Landauer{\textendash}Buttiker{\textquotedblright}
approach~\cite{landauer}, where the junction itself is
superconducting. We shall use this  formulation to study
the influence of the proximity effect on the transport properties
of a superconducting junction specifically for the case of two and
three \od {\textsl{interacting}} quantum wires
 and show how the simple case of
junction of a single \od \qw with a superconductor (\ns junction) is
different from the multiple wire junction counterpart. Because of
the existence of the \ar process, both electron and hole channels
take part in transport.  The power law dependence of the Andreev
conductance for the \ns junction case was first obtained using weak
interaction renormalization group (\wirgd) approach by {{Takane and
Koyama in Ref.~\onlinecite{jap1}}}. This was in agreement with
earlier results from bosonization~\cite{jap3}, which, however,
%
\begin{figure}[htb]
\begin{center}
\epsfig{figure=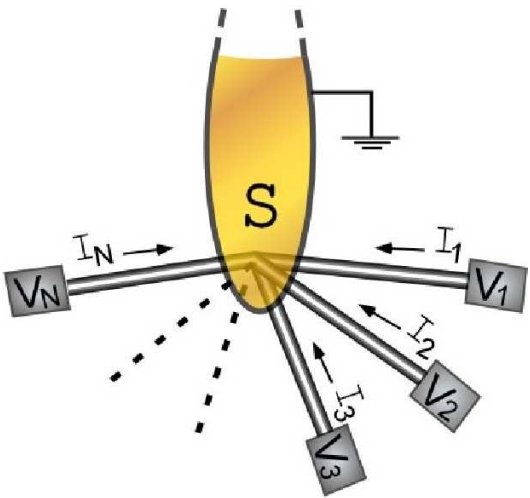,width=7cm}
\end{center}
\caption{Multiple wires connected to a superconducting junction.
The dashed lines represents the fact that the model can be
trivially extended to more than two wires.
{\textquoteleft}$a${\textquoteright} is the effective length of
the superconductor.} \label{nsnfig1}
\end{figure}
%
could  only handle perturbative analyses around the strong
back-scattering (\sbsd) and weak back-scattering (\wbsd) limits. The
\wirg approach, on the other hand,  can study the full cross-over
from \wbs limit to \sbs limit. Hence the \wirg approach is very
well-suited for studying problems where the aim is to look for non
trivial fixed points with intermediate transmissions and
reflections. This would be difficult using a bosonization approach.

For the \ns junction case, it was shown that the power law exponent for the
temperature dependence of conductance was twice as large
as the exponent for  a single barrier in a \qwd. This
happens because of the introduction of the extra hole channel due
to \ard. The \wirg approach takes into
account both {\textsl{(a)}} electron-electron interaction induced
forward scattering
processes which gives standard Luttinger liquid behavior, and
{\textsl{(b)}} electron-electron interaction induced
back-scattering processes which
give rise to non Luttinger liquid behavior (deviation from pure
power law behavior) as obtained by Matveev {\textsl{\etal~}} in
Refs.~\onlinecite{matveev} and \onlinecite{yue}.
The most interesting point to be noticed here is the fact that both
bosonization and \wirg give the same Luttinger liquid power law
dependence for the conductance of \ns junction although the \wirg
approach actually takes into account the extra process of
electron-electron back-scattering which usually leads to non
Luttinger liquid behavior. This happens because there is a
remarkable cancellation in perturbation theory which nullifies any
deviation from pure power law behavior of conductance.

In this article, we show that this kind of cancellation does not
happen for the \nsn junction or for that matter, for any junction
comprising of more than two \od quantum wires. Hence, deviations
from pure power law do exist. We show that due to the inter-play of
the proximity and the interaction effects, one gets a novel
non-monotonic behavior of conductance for the case of \nsn junction
as a function of the temperature. {\it Note that this is something
which cannot be obtained in a bosonization analysis which neglects
the back-scattering part of the electron-electron interaction.} We
extend our results to junctions with a ferromagnetic wire on one
side, \ie,
{\textsl{ferromagnet{\textendash}superconductor{\textendash}normal}}
(\fsnd) junctions and to junctions with ferromagnets on both sides,
\ie, {\textsl{ferromagnet{\textendash}superconductor{\textendash}
ferromagnet }} (\fsfd) junctions. Here we assume that the influence
of the junction between the superconductor and the ferromagnetic
wire has a very small effect on the spectrum of the superconductor
itself. Of course this is true only if the superconductor is large
enough. We also study transport through a superconductor at the
junction of three wires (and finally  extend it to ${\text{N}}$
wires). This generalizes the earlier work on
junctions~\cite{lal,das} to also include proximity effects.

In Sec.~\ref{sec:two}, we review the applicability and strength of
the \wirg approach when applied to
quantum impurity problems, such as, a normal junction of multiple
quantum wires, or a junction with a spin impurity. Then we discuss how one
can apply the \wirg technique to the problem studied in this paper.
In Sec.~\ref{sec:three}, we describe the set-up for our system,
\ie~ a superconductor at the junction of ${\text{N}}$ wires, in terms
of a scattering matrix and briefly discuss the symmetries of the
proposed model. We then perturbatively (in electron-electron
interaction strength) calculate the leading order logarithmic
corrections to both the normal reflection amplitude and the \ar amplitude
which, in turn, give corrections to the conductance via the
Landauer{\textendash}Buttiker formula.
In Sec.~\ref{sec:four}, we obtain the \rg equation for the \ns
junction and reproduce the known fixed points using our approach.
Then we derive the \rg equation for the symmetric \nsn junction
and obtain the \rg flow between various fixed points and analyse
the results of the study. Unlike, the renormalization group flow of the \ns
junction, which does not lead to any non-monotonicity, we show
that the inclusion of the \car and direct tunneling through the
superconductor gives rise to a non-monotonic conductance as a
function of the temperature. In Sec.~\ref{sec:five}, we present
our results for specific cases of \nsn.
In Sec.~\ref{sec:six}, we study the
three$-$wire$-$superconducting$-$junction and show the existence
of a fixed point which is analogous to the Griffith's fixed
point~\cite{lal,das} in the three$-$wire$-$normal$-$junction case.
Finally in Sec.~\ref{sec:seven}, we present our summary and discussions.

\section{\label{sec:two}\wirg vis-a-vis Bosonization}
Transport through a quantum scatterer (for instance, a simple
static barrier or a dynamical impurity like Kondo spin) in a \od
interacting electron gas is qualitatively different
from its higher dimensional
counterpart~\cite{k&f}. This is because, in \odd, due to
electron-electron interactions, the Fermi-liquid ground state
is destroyed and the electrons form a non-Fermi liquid ground
state known as as {\textsl{Luttinger
Liquid}}~\cite{haldane}.
The low energy dynamics of the \od system is governed mainly by
coherent particle-hole excitations around the left and the right
Fermi points. It is natural to use bosonic fields to
describe these low lying excitations.
This can be done by re-expressing the original fermions using boson
coherent state representation~\cite{usds,vondelft,solyom} which is
referred to as {\textsl{bosonization}}. But this approach only
allows for a perturbative analysis for transport around the limiting
cases of \sbs and \wbs for the quantum impurity problem. On the
other hand, if we start with a very weakly interacting electron gas,
it is possible to do a perturbative analysis in the
electron-electron interaction around the free fermion Hamiltonian,
but  treating  the strength of the quantum impurity
{\textsl{exactly}}. This allows us to study transport through the
impurity for any scattering strength. The strength of this approach
lies in the fact that even in presence of electron-electron
interaction, one can use single particle notions such as the
transmission and reflection amplitudes in order to characterize the
impurity. The idea is to calculate correction to transmission and
reflection amplitude perturbatively in the interaction strength.
Of course, since we are working in \odd, the perturbative
correction turns out to be logarithmically divergent. To obtain a
finite result, one has to sum up all such divergent contributions
to the transmission and reflection amplitudes to all relevant
orders at a given energy scale. This was first done by
Matveev {\textsl{
\etal~}} in Refs.~\onlinecite{matveev} and \onlinecite{yue}
in the context of a single (scalar) scatterer for both spin-less
and spin-full electrons using the ``poor man's scaling''
approach~\cite{anderson}. In the spin-less case, it was shown that
the logarithmic correction to the bare transmission amplitude (to
first order in interaction parameter parameterized by $\alpha$)
was
%
${{\delta T = 2\,\alpha \,T_0\, \left(1\,-\,T_0\right) \,
{\ln}(kd) }} $ and the explicit \rg equation for transmission probability
was $dT/dl=-2 \alpha T (1-T)$
%
where $k$ was the momentum of the fermion measured from $k_F$, $d$
was a short distance cut-off and $\alpha$ was the interaction
parameter given by
%
$\alpha =\alpha_1 - \alpha_2$ with $\alpha_1={{V(0)}/{2\pi \hbar
v_F}}$ and $\alpha_2={V(2k_F)} /{2\pi \hbar v_F} $.
%
The \rg equation  upon integration gave the transmission
probability as,
\beq T(L) = \dfrac{T_0  e^{-2 \alpha l}}{\left[{1 - T_0 + T_0
e^{-2\alpha l}}\right]}
   = \dfrac{T_0  \left(\frac{d}{L}\right)^{2\alpha}}
   {\left[1 - T_0 + T_0 \left(\frac{d}{L}\right)^{2\alpha}\right]}
\label{eq:one} \eeq
 Here, $l = -\ln(kd) = \ln({L/d})$ where $L$ is the length
scale. $l$ can also be measured as a function of the temperature by
introducing the thermal length, $L_T = ({\hbar v_F})/({k_B T})$.
$T_0$ is the bare transmission at the short distance cut-off, $d$.
It is easy to see from Eq.~\ref{eq:one} that for very small values
of $T_0$, $T_0$ can be neglected in the denominator of the
expression for $T(L)$ leading to a pure power law scaling consistent
with the power law known from bosonization in the \wbs limit.
Similarly for the spin-full electrons, it was shown that the
parameter $\alpha$ in the power law gets replaced by a new
parameter, $\beta$ given by $
 \beta = {(g_2-2g_1)}/{\pi \hbar v_F}$ where $g_2 = g_2(k)$ and $g_1 =
g_1(k)$ are momentum dependent functions or ``running coupling
constants". The momentum dependence of $\beta$ here is in sharp
contrast to the momentum independent $\alpha$ in the spin-less case.
At high momentum, (or equivalently, at the short distance cut-off
scale),
$g_1(d) = V(2k_F)$ and $g_2(d) = V(0)$.
Because of the extra logarithmic dependence coming from scaling of
the interaction parameter itself (see Eq.~\ref{intrg1} and
Eq.~\ref{intrg2} below), the expression for
transmission~\cite{matveev,yue}, no longer shows a pure power law
scaling even for small values of $T_0$. Instead $T(L)$ is now given by
\beq T(L) = \dfrac{\left[T_0\left[1+\alpha_1 \ln
\left(\frac{L}{d}\right)\right]^\frac{3}{2}
\left(\frac{d}{L}\right)^{(2\alpha_2-\alpha_1)}\right]}{\left[1 -
T_0 + T_0 \left[1 + 2\alpha_1 \ln
\left(\frac{L}{d}\right)\right]^\frac{3}{2}
\left(\frac{d}{L}\right)^{(2\alpha_2-\alpha_1)} \right] }
\label{spcondlog}
 \eeq
using the length scale dependence of $g_1(L)$ and $ g_2(L)$ given
by~\cite{solyom} \bea g_1(L) &=& \dfrac{V(2k_F)}{\left[1 +
\frac{V(2k_F)}{{\pi v_{F}}} \ln (\frac{L}{d})\right]}
\label{intrg1}
\\
g_2(L) &=& V(0) - \frac{1}{2}\, V(2k_F) +
\frac{1}{2}\,\dfrac{V(2k_F)}{\left[1\,+\,\frac{{V(2k_F)}}{{\pi
v_{F}}}\ln(\frac{L}{d})\right]} \non\\ \label{intrg2} \eea

%
%
%
 Note that in the
absence of electron-electron interaction induced back-scattering
(\ie,~when $V(2k_F)=0$), there is no correction to the power law
behavior. Hence, bosonization, which ignores electron-electron
back-scattering always results in power law behaviour. But, when
electron-electron interaction induced back-scattering is included,
the sign of $g_2-2g_1$ can change under \rg flow, and hence, there
can be a qualitative change in the behavior of the conductance. The
conductance actually develops a non-monotonic dependence on the
temperature; it first grows and then drops to zero.
But, except for this non-monotonic behavior of conductance for the
spin-full case, there is no new physics which is found by studying
the full crossover from \wbs to \sbsd. In conclusion, both
bosonization and \wirg methods predict that for the single scatterer
problem there are only two fixed points - {\textsl{(a)}} the
perfectly back-scattering (no transmission) case is the stable fixed
point and {\textsl{(b)}} the no back-scattering (perfect
transmission) case is the unstable fixed point. There are no fixed
points with intermediate transmission.

It was first shown by Lal {\textsl{\etal}} in Ref.~\onlinecite{lal},
using the \wirg approach that even though there are only two fixed
points for the two$-$wire$-$junction, surprisingly enough, the
three$-$wire$-$junction has a host of fixed points, some of which
are isolated fixed points while others are one parameter or multi
parameter families of fixed points. It was also shown to be true for
more than three wires. From this point of view, the physics of a
two$-$wire$-$junction is different from its
three$-$wire$-$counterpart. The three$-$wire$-$junction was also
studied using bosonization and conformal field theory
methods~\cite{nayak,affleck1,affleck2,das2}, which confirmed some of
the fixed points found using \wirgd. It also  gave some extra fixed
points which were related to charge fractionalisation at the
junction, and which could not be seen within the \wirg approach.
The \wirg method was further extended to complicated systems made
out of junctions of \qw which can host resonances and
anti-resonances in Ref.~\onlinecite{das}. The scaling of the
resonances and anti-resonances were studied for various
geometries which included the ring and the stub geometry. In
particular, it was shown that for a multiple$-$wire$-$junction,
the \rg equations for the full $S$-matrix characterizing the
junction take a very convenient matrix form,
\beq \dfrac{dS}{dl} = - (S F^\dagger S ~-~ F) \label{eqlal}
\eeq
where $S$ is the scattering matrix at the junction and $F$ is a
diagonal matrix that depends on the interaction strengths and the
reflection amplitude in each wire. The advantage of writing the \rg
equation this way is that it immediately facilitates the hunt for
various fixed points. All one needs to do is to set the matrix on
the LHS of Eq.~\ref{eqlal} to zero. This will formally provide us
with all the fixed points associated with a  given $S$-matrix. This
approach was further extended in Refs.~\onlinecite{ravi1,ravi2} to
study the multiple$-$wire$-$junction with a dynamical scatterer,
\ie~ a (Kondo) spin degree of freedom. The coupled \rg equations
involving the Kondo couplings, $J_{ij}$ as well as the $S$-matrices
were solved. For different starting scalar $S$-matrices, the \rg
flows of the Kondo couplings was studied. The temperature dependence
of the conductances was shown to have an interesting interplay of
the Kondo power laws as well as the interaction dependent power
laws. Finally, the \wirg method was also extended to the case of \ns
junction~\cite{jap1,jap3}. In the vicinity of the superconductor, it
is well-known that the system is described by holes as well as
electrons~\cite{degennes}. Hence the $S$-matrix characterizing the
junction not only includes the electron channel but also the hole
channel. Naively, one might expect that in the presence of
particle-hole symmetry, the only effect of including the hole
channel would be to multiply the conductance by a factor of two (in
analogy with inclusion of spin and imposing spin up-spin down
symmetry). However, it was shown~\cite{jap1,jap3} that in the
vicinity of a superconductor, the proximity induced scattering
potential that exists between electron and holes, also gets
renormalized by electron-electron interactions. When this scattering
is also taken into account, the correction to the scattering
amplitude to first order in the interaction parameter depends on
$(2g_2-g_1)$ instead of $(g_2-2g_1)$. It is worth stressing that
this particular linear combination of the interaction parameters
($g_i$'s) is independent of the scaling as the logarithmic factors
($l = \ln(kd)$) in Eqs.~\ref{intrg1} and \ref{intrg2} cancel each
other. Hence, there is no non-monotonic behavior of the conductance
in this case. The \wirg predicted only two fixed points, the Andreev
fixed point (perfect \ard) which turns out to be an unstable one and
the perfectly reflecting fixed point which is the stable fixed
point. The \ns junction has also been studied using
bosonization\cite{jap2}. It is easy to check that the power laws
resulting from bosonization agree with those obtained from the
\wirg, when the electron-electron interaction induced
back-scattering ( which is dropped in the bosonization method) is
ignored.

In this article, we apply the \wirg method to the superconducting
junction of multiple quantum wires. We note that we now have two
complications - {\textsl{(a)}} multiple wires are connected to the
junction and {\textsl{(b)}} we have both electron and hole
channels connected to the junction. So in this case, even for the
\ns junction, we have two spin channels as well as the electron
and hole channels, so the scattering matrix is four component. For
$N$ wires, the scattering matrix is $4N \times 4N$-dimensional.
Although, we expect our method to work well even in this case,
there is one caveat we must keep in mind. We have incorporated the
effect of the superconductor as a boundary condition on the \qw
and neglected any internal dynamics of the superconductor itself.
This should work reasonably well as long as we are studying
transport at energies much below the superconducting gap. Our main
result here is that the conductance across the junction depends on
both $g_1$ and $g_2$ and not on a a special combination $2g_2-g_1$
(as in \ns case) which does not get
renormalized under \rg flow.%
 Hence, the cancellation of the logarithmic terms in the effective
interaction parameter is specific to the \ns case and is not true
in general. For $N \ge 2$ wires attached to a
superconductor, we expect a non-monotonic form of the conductance.
We also expect to get a
host of fixed points with intermediate transmission and
reflection, knowledge of which can be of direct relevance for
application to device fabrication of such geometries.

\section{\label{sec:three} Superconducting junction with $N$ quantum wires}
Let us consider multiple ($N$) quantum wires meeting at a junction on which
a superconducting material is deposited as depicted in
Fig.~\ref{nsnfig1}. The wires are parameterized by coordinates
$x_i$, with the superconducting junction assumed to be at $x_i=0$.
We consider a situation where the effective width `$a$' of the
superconductor between two consecutive wires is of the order of
the phase coherence length of the superconductor (size of the
Cooper pair). For our purpose, it is safe to ignore the
finiteness of `$a$' and effectively treat the junction of \qw as a
single point in space with an appropriate boundary condition.
We parameterize the junction by the following quantum mechanical
amplitudes. There are two kinds of reflection amplitudes: the
normal reflection amplitude ($r_{ii\,s_i\,s_i}$) and the \ar
amplitude ($r_{A{ii}\,s_i\,s_i}$) on each wire.
In addition, there are two kinds of transmission amplitudes between
different wire: the co-tunnelling (\ctd) amplitude ($t_{ij \,s_i
\,s_j}$) and the \car amplitude ($t_{Aij \,s_i\,s_j}$). The indices
$s_i,s_j$ refer to the spin of incoming and outgoing particles. As
we consider a singlet superconductor at the junction, spin remains
conserved in all the processes mentioned above.
Thus, the boundary conditions are parametrized by a
$4N \times 4N$ scattering matrix for $N$ quantum wires
connected to a superconducting junction.

Let us now consider the various symmetries that can be imposed to
simplify the $4N \times 4N$ matrix. We impose
%
%
particle-hole symmetry, \ie, we assume that the reflection and
transmissions are the same for particles (electrons) and holes.
Further, in the absence of a magnetic field, spin symmetry is
conserved which implies that the various transmission and
reflection amplitudes for spin up-down electrons and holes are
equal.
%
(This symmetry breaks down in the presence of magnetic fields, or
in the case of ferromagnetic wires). Also, since we assume
that all the wires, connected to the superconductor,
are indistinguishable, we can impose a wire
index symmetry. (This symmetry again can be broken if we take some
ferromagnetic and some normal wires attached to the
superconductor). On imposing these symmetries, the $S$-matrix for
a two-wire system is given by
\beq S \quad = \quad \begin{vmatrix} ~S_{\up} & 0~ \\
~0 & S_{\dn}~
\end{vmatrix}
\non \eeq \noindent with \beq
 S_{\up} \quad =\quad   S_{\dn} \quad
 = \quad \begin{vmatrix} ~r & t & r_A & t_A  ~\\
~t & r & t_A & r_{A} ~\\
~r_{A} & t_A & r & t ~\\
~t_A & r_{A} & t & r ~
\end{vmatrix}
\label{smat1} \eeq Here $r$ stands for normal reflection of
electron or hole in each wire, and $r_{A}$ represents \ar from
electron to hole or vice-versa in each wire. $t$ represents
the elastic \ct amplitude ($t = t_{12} = t_{21}$) while $t_{A}$
represents the \car amplitude ($t_{A} = t_{A12} = t_{A21}$). For
the spin symmetric case, there are two such matrices, one for spin
up electrons and holes and one for spin down electrons and holes.
Note that this is the relevant $S$-matrix at energy scales
(temperature and applied voltage on the wires) $k_B \,T, e V_i \ll
\Delta$, where $\Delta$ is the superconducting gap energy.
The competition between \ct and \car has been analysed
before~\cite{hekking1,hekking2} and also different ways of
separating the contributions experimentally
have been considered~\cite{beckmann}.
However, the effect of electron-electron
interactions within the wires has not been considered for the \nsn
case. It is worth emphasizing here that if such \nsn junctions are
made out of \od systems like carbon nanotubes, then the effect of
electron-electron interactions can influence the low energy
dynamics significantly.

The Landauer$-$Buttiker conductance matrix for the \nsn case
can be written, in the regime where $k_B \,T, eV_i \ll \Delta$,
as~\cite{hekking1}
\beq
\begin{vmatrix} ~I_1~ \\
~I_2~ \\
\end{vmatrix}
~=~
\begin{vmatrix} ~
G_{A}+G_{CA}+G_{CT}& G_{CA}-G_{CT}~ \\
~G_{CA}-G_{CT}& G_{A}+G_{CA}+G_{CT}~
\end{vmatrix}
\begin{vmatrix}
~V_1~ \\
~V_2~
\end{vmatrix}
\eeq The conductances here are related to the elements of the
$S$-matrix: $G_{A} \propto |r_{A}|^2$, $G_{CT} \propto |t|^2$ and
$G_{CA} \propto |t_{A}|^2$. $G_{A}$ is the conductance due to the
\ar that occurs at a single \ns junction, whereas $G_{CT}$ and
$G_{CA}$ are the conductance due to the elastic \ct and \car
processes respectively, both of which involve transmissions
between two wires and give contributions with opposite signs to
the sub-gap conductance between the two wires, $G_{CA}-G_{CT}$ .
The generalization of this to $N > 2$ is straightforward, and some
details are presented in Sec.~\ref{sec:six}.

\section{\label{sec:four} \wirg study of junctions}

We study the effects of inter-electron interactions on the
$S$-matrix using the renormalization group (\rgd) method
introduced in Ref.~\onlinecite{yue}, and the generalizations to
multiple wires in Refs.~\onlinecite{lal,das}. The basic idea of
the method is as follows. The presence of back-scattering
(reflection) induces Friedel oscillations in the density of
non-interacting electrons. Within a mean field picture for the weakly
interacting electron gas, the electron not only scatters off the
potential barrier but also scatters off these density oscillations
with an amplitude proportional to the interaction strength. Hence
by calculating the total reflection amplitude due to scattering
from the scalar scatterer and from the Friedel oscillations
created by the scatterer, we can include the effect of
electron-electron interaction in calculating transport.
This can now be generalized to the case where there is, besides
non-zero reflection, also non-zero \ar.

To derive the \rg equations in the presence of Andreev processes,
we will follow a similar procedure to the one followed in
Ref.~\onlinecite{lal}. The fermion fields on each wire can be written
as,
\beq \psi_{is} (x) = \Psi_{I\,is}(x)\, e^{i\,k_F\,x} \,+\,
\Psi_{O\,is}(x)\, e^{-i\,k_F\,x} \eeq
where $i$ is the wire index, $s$ is the spin index which can be
$\up,\dn$ and $I,O$ stands for outgoing or incoming fields. Note
that $\Psi_{I}(x) (\Psi_O (x))$ are slowly varying fields on the
scale of $k_F^{-1}$ and contain the annihilation operators as well
as the slowly varying wave-functions. For a momentum in the
vicinity of $k_F$, the incoming and outgoing fields (with
the incoming field on the $i^{th}$ wire) can be Fourier
expanded in a complete set of states and the electron field  can be
written as
\bea
\Psi_{is}(x) &=& \int ~dk~ \Big[\, b_{ks} \,\epp \,+\,
d_{ks}^\dagger \, \epm  \non\\
 &+& r \,b_{ks}\, \emp  \,+
 \, r^\star \,d_{ks}^\dagger \, \emm  \non\\
 &+&  r_A \,d_{ks}\, \emm  \,+\, r_A^\star \,b_{ks}^\dagger\,
 \emp \Big]
\non\\
\Psi_{(j\ne i)s}(x) &=& \int ~dk~ \Big[ tb_{ks} \,\epp \,+\,
t d_{ks}^\dagger \, \epm  \non\\
 &+&  t_A \,d_{ks}\, \emm  \,+\, t_A^\star \,b_{ks}^\dagger\,
 \emp \Big]
\non\\
\eea
where $b_{ks}$ is the electron destruction operator and $d_{ks}$ is
the hole destruction operator. Note that we have chosen to quantize
the fermions in the basis of the space of solutions of the Dirac
equation, in the presence of a potential which allows for normal as
well as Andreev scattering. We have also allowed for both incident
electrons and holes. We find that  (dropping a constant background
density),
\bea
\scxone{\rho_{is}(x)} = \scxone{\Psi_{is}^\dagger \Psi_{is}^{}} =
\frac{i}{4 \pi x}~ \Big[(r^\star e^{2ik_{F}x} - r e^{-2ik_{F}x})
\non\\ ~+~ (r e^{2ik_{F}x}-r^\star e^{-2ik_{F}x})\Big] \non
\\
\label{density} \eea where the two terms corresponds to the
density for electrons and holes respectively. Here we have also
used the fact that due to the proximity of the superconductor, the
amplitude to create (destroy) a spin $s$ electron and destroy
(create) a spin $s$ hole is non-zero {\textemdash} \ie, the
Boguliobov amplitudes $\scxone{d_{k-s}^\dagger b_{ks}^\dagger} =1=
\scxone{b_{ks}d_{k-s}}$, besides the normal amplitudes
$\scxone{d_{ks}^\dagger d_{ks}}= \scxone{b_{ks}b_{ks}^\dagger}
=1$. (This is of course true only close to the superconductor. We
have checked that this gives the same result as solving the
Boguliobov{\textendash}de Gennes equation as done in
Ref.~\onlinecite{jap3}) Hence, besides the density, the
expectation values for the pair amplitudes
$\scxone{\Psi_{is}^\dagger\Psi_{is}^\dagger}$ and its complex
conjugate $\scxone{\Psi_{is}\Psi_{is}}$ are also non-zero and are
given by (dropping the wire index)
\bea \scxone{\psi_{O\,\up}^\dagger \psi_{I\,\dn}^\dagger} ~=~ -
\scxone{\psi_{O \,\dn}^\dagger \psi_{I\,\up}^\dagger} ~=~ \frac
{-i\, r_{A}}{4 \pi x}
  \non \\\And \quad
\scxone{\psi_{O\,\up} \psi_{I \,\dn}} ~=~ - \scxone{\psi_{O\,\dn}
\psi_{I\,\up}} ~=~ \frac {-i \,r_{A}^\star} {4 \pi x}. \label{boguliobov}
\eea
So, we see that the Boguliobov amplitudes fall off as $1/x$
just like the normal density amplitudes.

We now allow for short-range density-density interactions between
the fermions
\beq
\hmi = \frac{1}{2} \, \int dx \,dy \, \left(\sum_{s\,=\,\up,\dn}
\rho_{s}\right) \,V(x-y)\,\left(\sum_{s\,=\,\up,\dn}
\rho_{s}\right)
\eeq
to obtain the standard four-fermion interaction Hamiltonian for
spin-full fermions as
\bea
\hmi & = & \int dx \Big[ g_1 \Big(\psiiudg  \psioudg \psiiu \psiou
\,+\, \psiiddg \psioddg \psiid \psiod
\nonumber\\
&+&   \psiiudg \psioddg \psiid \psiou \,+\, \psiiddg \psioudg \psiiu
\psiod\Big)
\nonumber \\
&+&  g_2  \Big(\psiiudg \psioudg \psiou \psiiu
+ \psiiddg \psioddg  \psiod \psiid
\nonumber \\
&+&  \psiiudg \psioddg \psiod \psiiu  \,+\,
\psiiddg \psioudg \psiou  \psiid\Big)\Big]
\nonumber \\
&& \eea where $g_1$ and $g_2$ are the running coupling constants
defined in Sec.~\ref{sec:two} (Eq.~\ref{intrg1} and
Eq.~\ref{intrg2}).

Using the expectation values for the fermion operators, the
effective Hamiltonian can be derived using a Hartree$-$Fock (\hfd)
decomposition of the interaction. The charge conserving \hf
decomposition can be derived using the expectation values in
Eq.~\ref{density} and leads to the interaction Hamiltonian
(normal) of the following form on each half wire,
\bea
\hmi^N &=& \dfrac{-i(g_2-2g_1)}{4\pi}  \int_0^\infty \dfrac{dx}{x}
\Big[r^\star \left(\psiiudg \psiou + \psiiddg \psiod \right)
\nonumber \\
&-& r \left(\psioudg \psiiu + \psioddg \psiid \right)\Big]. \eea
(We have assumed spin-symmetry \ie~ $r_{\up} = r_{\dn} = r$.) This
has been derived earlier~\cite{lal}. Using the same method, but
now also allowing for a charge non-conserving \hf decomposition
with the expectation values in Eq.~\ref{boguliobov}, we get the
(Andreev) Hamiltonian \bea
\hmi^A &=& \dfrac{-i(g_1+g_2)}{4\pi}  \int_0^\infty  \dfrac
{dx}{x} \Big[-r_{A}^{\star}\big(\psiiudg \psioddg +
 \nonumber \\
&& \psioudg \psiiddg \big)
 + r_{A}\left( \psiod\psiiu + \psiid\psiou \right) \Big].\eea Note
that although this
appears to be charge non-conserving, charge conservation is taken
care of by the $2 e$ charge that flows into the superconductor
every time there is an Andreev process taking place.

The amplitude to go from an incoming electron wave to an outgoing
electron wave under $e^{-i{\hmi^N}t}$ (for electrons with spin)
was derived in Ref.~\onlinecite{lal} and is
given by
\beq {-\alpha \, r_{s} \over 2} \, \ln (kd) \label{Friedeln} \eeq
where $\alpha = (g_2-2g_1) / 2\pi\hbar v_F $ and $d$ was a short
distance cut-off. Analogously, the amplitude to go from an
incoming electron ${\sf{e_{in}}}$ wave to an outgoing hole
${\sf{h_{out}}}$ wave under $e^{-i{\hmi^A} t}$ is given by \bea
e^{-i\,\hmi^A \, t}
\ket{{\mathsf{e_{in}}},s,k}
\hskip -2.5 cm
\non \\
&=& -i \int
\dfrac{dk^\prime}{2\pi}\Bigg[\ket{{\mathsf{h_{out}}},s^\prime,k^\prime}
 \me{\hmi^A}
{{\mathsf{h_{out}}},s^\prime,k^\prime}{{\mathsf{e_{in}}},s,k}\Bigg]
\hskip -0.5 cm
\non \\
&=& \dfrac{-i(g_1\,+\,g_2)\, r_A} {4 \, \pi \, \hbar \, v_F}
\int_{}^{} \dfrac{dx}{x}  e^{-2\,i\,k\,x} \,
\ket{{\mathsf{h_{out}}},s^\prime,k^\prime}  \eea
where $s \ne s^\prime$. Hence, the amplitude for an incoming
electron to be scattered to an outgoing hole is given by
\beq {\alpha^\prime \,r_{A} \over 2} \, {\ln (kd)}
\label{FriedelA} \eeq
where $\alpha^\prime = (g_1+g_2)/ 2 \pi  \hbar  v_F $. Note also
that $\alpha$ and $\alpha'$ are themselves momentum dependent, since
the $g_i$'s are momentum dependent. The amplitude for an
incoming electron to go to an outgoing electron on the same wire
is governed by the interaction parameter $\alpha =(g_2-2g_1) /
2\pi\hbar v_F $ which has the possibility of chaging sign under
RG evolution, because of the relative sign between $g_1$ and $g_2$.
On the other hand, $\alpha^\prime = (g_2+g_1) / 2\pi\hbar v_F $ can
never change its sign.

\subsection{\ns Junction}\label{ss:wirg_ns}
 The amplitudes in Eqs.~\ref{Friedeln} and \ref{FriedelA} are
corrections to the reflections of electrons from Friedel oscillations and
from the pair potential respectively.
We can combine them with the $S$-matrix at the junction to find
the corrections to the amplitudes of the $S$-matrix. For an \ns
junction, there is only one wire coupled to the superconductor and
the $S$-matrix is just $2 \times 2$ for each value of the spin and is
given by
\beq S \quad=\quad \begin{vmatrix} ~r &  r_{A}~ \\
~r_{A}& r~
\end{vmatrix}
\label{smat2} \eeq
Here $r$ is the normal refelction amplitude and $r_A$ is the
Andreev reflection amplitude. So we only need to compute the
corrections to $r$ and $r_A$ in this case.

\begin{figure}[htb]
\begin{center}
\epsfxsize=8.0cm \epsfysize=8.0cm
\epsfig{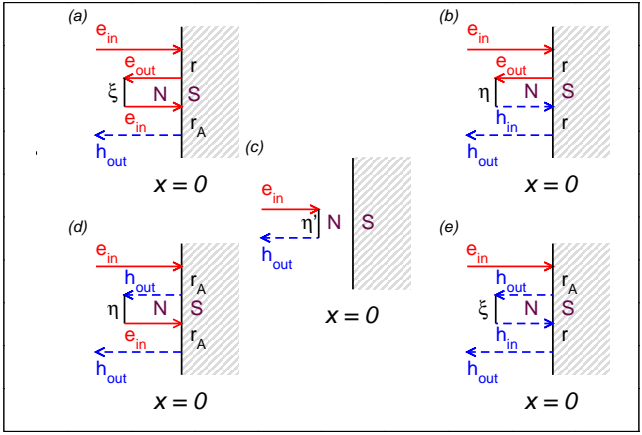}
\end{center}
\caption{{The processes that contribute to the amplitude for an
incoming electron to transform to an outgoing hole. Note that all
the processes shown here are to first order in the interaction
parameters since they only involve a single scattering from a
Friedel oscillation or the pair potential. Process $(c)$ involves
scattering from a pair potential before the electron reaches the
junction. The remaining processes involve two reflections from the
junction and a scattering from the Friedel oscillation or the pair
potential. In the diagrams, $\xi = \frac{1}{2}\,\alpha
\,r^{\star}\, {\ln (kd)}$, $\eta = -\,\frac{1}{2}\,\alpha^\prime
\, r_{A}^{\star}\, \ln (kd)$ and $\eta^\prime=\frac{1}{2} \,
\alpha^\prime \, r_{A} \, \ln (kd)$.}} \label{nsnfig2}
\end{figure}

We find that there are five processes which contribute to the
amplitude $r_A$ to first order in the interaction parameter.
These are illustrated in Fig.~\ref{nsnfig2}.

\vskip +1.3cm
%
%
Adding all the contributions, we obtain the change in the \ar
amplitude $r_{A}$ that takes an incoming electron to an outgoing
hole given by
\bea \delta r_{A} &=&  \dfrac{\alpha^\prime}{2}  \big[\,r_{A}\,-\,
r_A^{\star} \left( r^{2} + r_{A}^{2} \right)\,\big]
 \, \ln(kd)  \non\\
&& +~ \alpha \, |r|^2 \,  r_{A} \, \ln(kd) \label{rans} \eea
%
in agreement with Ref.~\onlinecite{jap2}. For an incoming electron
reflected back as an electron, we find the small correction in the
amplitude $\delta r$ given by~\cite{yue,lal}
\bea {{\delta r}} &=& {{-\alpha^\prime \, |r_A|^2 \, r \, \ln(kd)
}}
\nonumber \\
&& {{+~ \frac{\alpha}{2} \, \left[\, r_A^2 \, r^{\star} \,-\,
r\left(1\,-\,|r|^2\right)\, \right] \ln(kd) }} \label{ransnew0}
\eea
We replace $-\ln(kd)$ by $dl$ using the ``poor man's scaling"
approach~\cite{anderson} to obtain the
 \rg equation for $r_A$ as
  \bea \frac {dr_{A}}{dl} &=&
 -\,\frac{\alpha^\prime}{2} \left[r_{A}
\,-\, r_A^{\star}\left(r^{2} \,+\, r_{A}^{2}  \right)
\right]
\,-\, \alpha \, |r|^2   r_{A}\label{ransnew1} \eea
Using the unitarity of  the $S$-matrix ($\vert \, r_A \, \vert ^2
+ \vert\, r^2 \, \vert = 1$ and $r_A^\star  r + r_A r^\star = 0
$), we can simplify the RHS of the above equation to obtain

  \bea \frac {dr_{A}}{dl} &=&
  -\, \left(\,\alpha \, + \, \alpha^\prime \,\right) \, r_A
  \left(\,1 - \vert r_A \vert ^2 \,
\right)  \label{ransnew2} \eea
 Note that the combination $\alpha
\,+\,\alpha^\prime = (2\,g_2 - g_1)/2\pi \hbar v_F$ which appears
in the \rg equation does not flow under \rgd. This can be
seen from Eqs.~\ref{intrg1} and \ref{intrg2} which shows that
$ (2\,g_2 - g_1)/2\pi \hbar v_F = (2V(0) -V(2k_F))/2\pi\hbar v_F$.
This means that $r$ and $r_A$ either monotonically increase or
decrease as a power law depending on the sign of
$\alpha\,+\,\alpha^\prime$. From Eq.~\ref{ransnew2}, we also
observe that $\vert r_A \vert = 0$ and $ \vert r_A \vert = 1$
correspond to the insulating and the Andreev fixed points of the
\ns junction respectively. One can easily see from the \rg
equations that $\vert r_A \vert = 0$ is a stable fixed point and
$\vert r_A \vert = 1$ is an unstable fixed point.

\subsection{\nsn Junction} \label{ss:wirg_nsn}
\begin{figure}[b]
\hskip -4.2cm
\includegraphics[width=4.2cm,height=3.5cm]{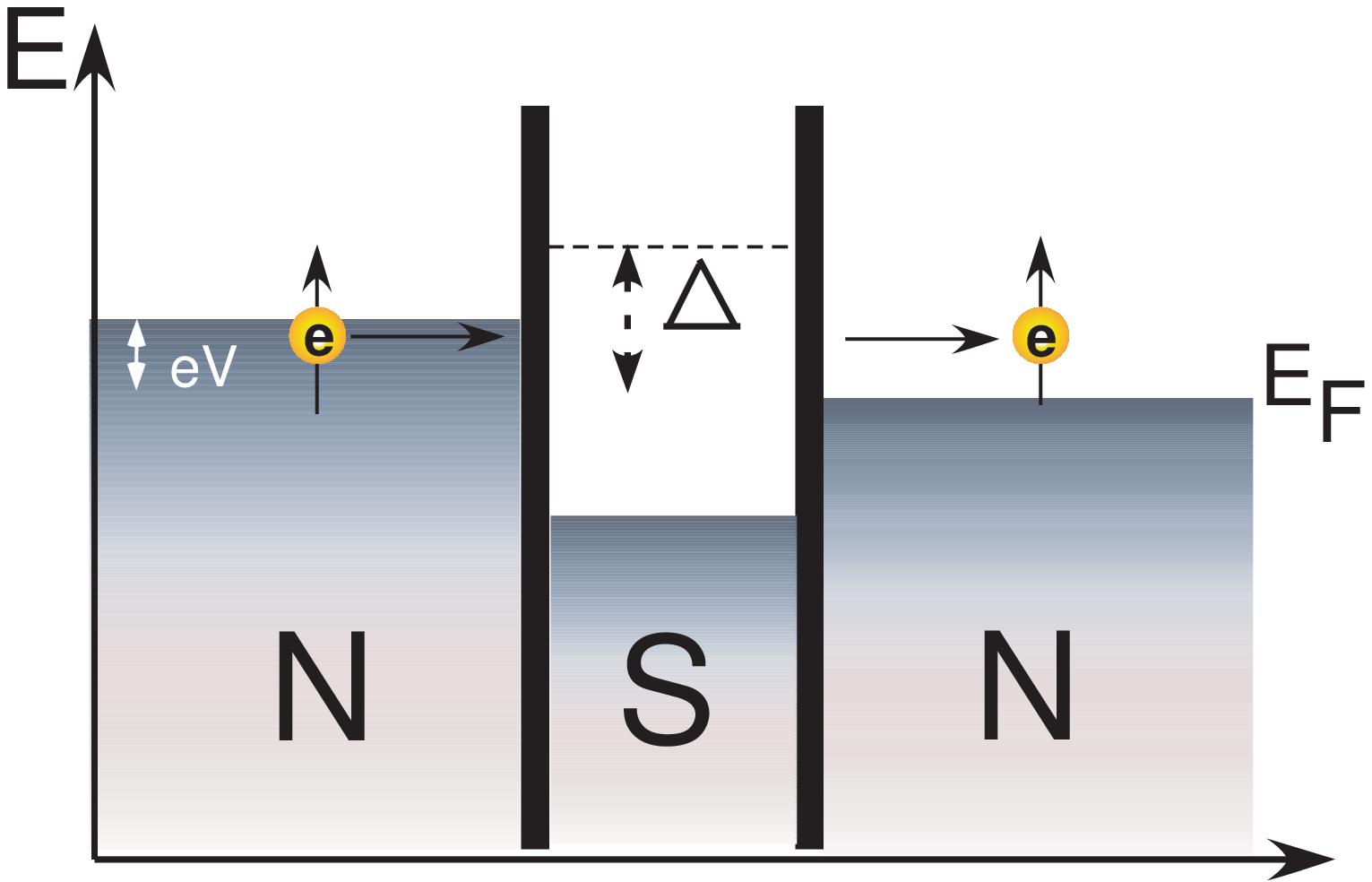}
 \vskip -3.5cm \hskip 4cm
\includegraphics[width=4.2cm,height=3.5cm]{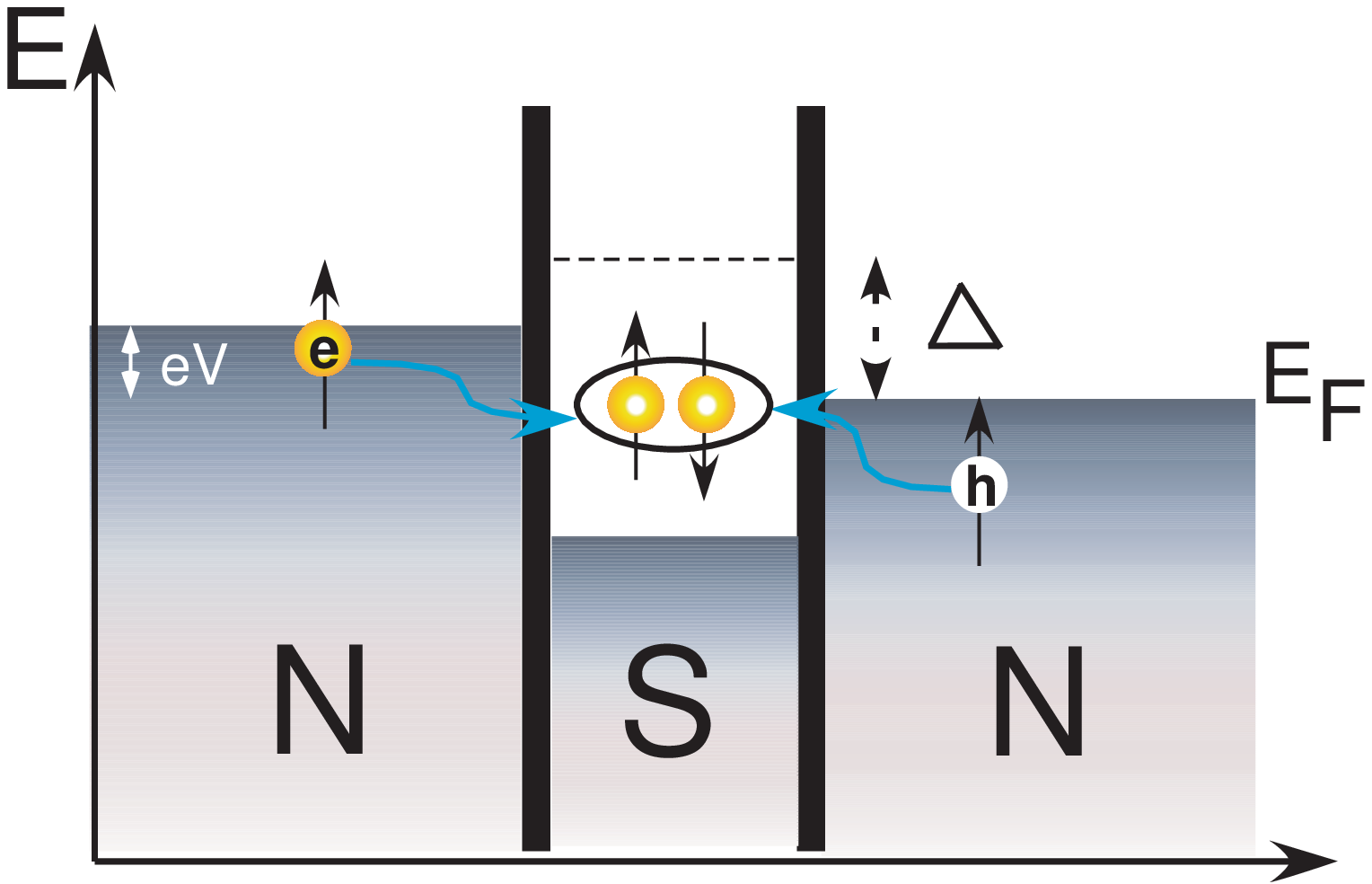}
 \caption{{Electron \ct  with bare
amplitude $t$ is shown in the left plot and \car with bare amplitude $t_A$
is shown in the right plot.}}
\label{nsnfig3}
\end{figure}

\begin{figure}[htb]
\begin{center}
\epsfxsize=8.0cm \epsfysize=8.0cm
\epsfig{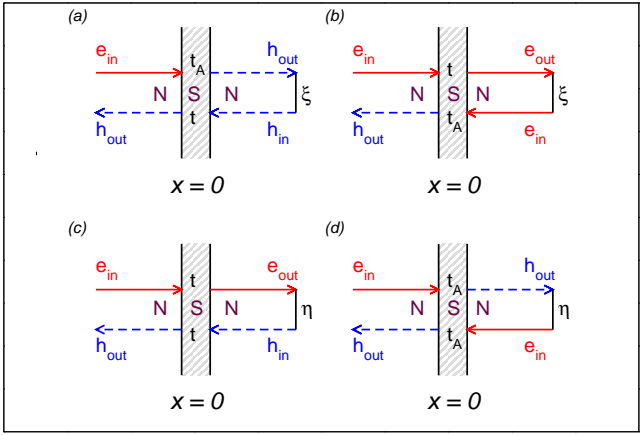}
\end{center}
\caption{The extra processes that contribute to the amplitude for
an incoming electron to transform to an outgoing hole on the same
wire, due to the second wire. Processes (a) and (b) are
transmitted to the second wire and reflected by the Friedel
oscillation whereas (c) and (d) are transmitted to the second wire
and reflected by the pair potential.} \label{nsnfig4}
\end{figure}

In this subsection, we shall consider an \nsn junction. Here in
addition to the two reflection channels, we also have two channels
for transmission - the direct transmission of an electron to an electron
through \ct process and the transmission of an electron to a hole via
\card. These processes are depicted in Fig.~\ref{nsnfig3}. The $S$-matrix at
the junction is $8 \times 8$ in this case as given in
Eq.~\ref{smat1}.
The number of processes that contribute in this case is thirty
four, since we also need to include terms that transmit electrons
or holes through the junction. For instance, for the
renormalization of the \ar term, besides the terms corresponding
to the \ns junction, we also have to include processes in which
the electron is incident from wire $1$, goes through the junction
to wire $2$, Andreev reflects from the pair potential on wire $2$
and then comes back through the junction, as shown pictorially in
 Fig.~\ref{nsnfig4}(c).

Collecting all the nine processes that contribute to first order
in $\alpha$ and $\alpha'$ to the reflection amplitude, we find
that \bea \dfrac{dr}{dl} &=& - \bigg[\dfrac{\alpha}{2} \, \left[
(t^2\,+\,r_{A}^2\,+\,t_{A}^2)\,r^\star\,-\,r(1-|r|^2)\right]
\non\\
&-& \alpha'\, ( r \,|r_{A}|^2 \,+\, r_{A}^\star \,t_{A}\,t) \bigg]
\label{rnsn} \eea Similarly, adding  up the contributions from the
 nine processes that contribute to $r_A$, we find that
\bea \dfrac{dr_{A}}{dl} &=&
-\bigg[\alpha(|r|^2\,r_{A}\,+\,t\,t_{A}\,r^\star)
\non\\
&+&
\dfrac{\alpha'}{2}(r_{A}\,-\,(r^2\,+\,r_{A}^2\,+\,t^2\,
+\,t_{A}^2)\,r_{A}^\star)
\bigg] \label{ransn} \eea
\vskip +1.3cm
%
Moreover, here besides the reflection parameters, we also need to
compute the renormalizations of the transmissions to first order
in $\alpha$ and $\alpha^\prime$. The \rg equations for $t$ and
$t_A$ are also obtained by considering all possible processes that
ultimately have one incoming electron and one outgoing electron
(for $t$) and one incoming electron and one outgoing hole (for
$t_A$) and are either reflected once from the Friedel potential or
the pair potential. They are found to be
\bea \dfrac{dt}{dl} &=&
-\,\bigg[ \alpha \, ( |r|^2\,t \,+\, r^\star \,r_A \,t_A) \non\\
&-&  \alpha^\prime\,(|r_A|^2 \,t \,+\, r \,r_A^\star\, t_A) \bigg]
\label{tnsn}\\
\dfrac{dt_A}{dl} &=& -\,\bigg[\alpha (r^\star\, r_A\,t \,+\,
|r|^2\,
t_A) \non\\
& -& \alpha^\prime\,(r\,t\,r_A^\star \,+\,|r_A|^2\,t_A)\bigg]
\label{tansn} \eea

Just as was done for the normal junction (Eq.\ref{eqlal}),
we can express the \rg
equations for the superconducting junction
in a compact matrix form~\cite{lal},
\beq \frac{dS}{dl} = {\tilde F} - S {\tilde F}^\dagger S \eeq
where the matrix $S$ is given in Eq.\ref{smat1} and
$\tilde F$ depends on the interaction parameters $\alpha
= (g_2-2g_1)/2\pi \hbar v_F$ and $\alpha^\prime = (g_1+g_2)/2\pi
\hbar v_F$. $\tilde F$ is {\textsl{non-diagonal}} matrix (unlike the case
in Ref.~\onlinecite{lal}) and is given by
\beq {\tilde F} ~=~
\begin{vmatrix}
~\frac{\alpha r}{2}& 0 & \frac{-\alpha^\prime r_{A}} {2} &0~  \\
~0& \frac{\alpha r}{2}& 0& \frac{-\alpha^\prime r_{A}}{2} ~ \\
~\frac{-\alpha^\prime r_{A}}{2} &0 & \frac{\alpha r}{2}& 0 ~\\
~0& \frac{-\alpha^\prime r_{A}}{2} & 0 & \frac{\alpha r}{2}~ \\
 \end{vmatrix}~.
\label{fmat} \eeq
It is easy to check that all the \rg equations
are reproduced from the matrix equation. The matrix form
also makes the generalization to $N$ wires case notationally
simple and makes the search for various fixed point much easier.
This will be discussed in the last section. But note that these
equations have to be augmented by Eqs.~\ref{intrg1} and
\ref{intrg2} to get the full set of \rg equations.

Let us now look at some of the fixed points of the $S$-matrix.
Clearly, the fixed points occur when $F-SF^\dagger S=0$ or when
$FS^\dagger$ is hermitian. There are several possibilities and we
list below some of them.

\begin{enumerate}
\item[]{{\textsl{Case I:}} Any one of the four parameters is non-zero}\\
{\textsl{(a)}} $t=1$, $r=r_{A}=t_A=0$, fully transmitting fixed point (\tfpd)\\
{\textsl{(b)}} $r=1$, $r_{A}=t_A=t=0$ fully reflecting fixed point (\rfpd)\\
{\textsl{(c)}} $r_A=1$, $r=t=t_A=0$, fully Andreev reflecting
fixed point
(\afpd)\\
{\textsl{(d)}} $t_A=1$, $r=t=r_A=0$, fully crossed Andreev
reflecting fixed point. (\cafpd)
\item[]{{\textsl{Case II:}} Any two are non-zero}\\
When both $r$ and $r_{A}$ are zero, the RHS of the \rg equations
identically vanishes as both the Friedel oscillation amplitude as
well as the pair potential amplitude in the wire become zero.
Hence any value of $t$ and $t_A$ remains unrenormalized  under
\rgd.
\item[]{{\textsl{Case III:}} Any three of them are non-zero}\\
We did not find any fixed point of this type.
\item[]{{\textsl{Case IV:}} All four of them are non-zero}\\
Here, we get a fixed point when $r_1=r_2=t=t_A=1/2$ and
$r_{A1}=r_{A2}=-1/2$. This is the most symmetric $S$-matrix
possible for the \nsn case. Since it is a symmetry-dictated fixed
point with intermediate  transmission and reflection, we shall
refer to it as symmetric fixed point (\sfpd).
\end{enumerate}
We will study the \rg flows  near some of these fixed points in
the next section.

\subsection{\fsd, \fsf and \fsn Junctions}
\label{ss:wirg_ferro}

We can also consider junctions where one or more of the wires are
spin-polarised, with Fermi distributions for the spin up and
down electrons being different.
As long as at least one of the wires is ferromagnetic, the spin up-spin
down symmetry of the system is broken. This means that we can no
longer impose $S_\up = S_\dn$ on the $S$-matrix parametrising the scattering
as we had in  Eq.\ref{smat1}. We now need to choose an $S$-matrix
with indices  $\uparrow$ and $\downarrow$ denoting the spin.
For the \fsn case (and the \fsf case where the ferromagnets on the
two sides are not identically polarized) the wire index
symmetry is also broken. Hence, the $S$-matrix chosen must also break
the wire-index symmetry. Note that for the ferromagnetic wire, the
amplitude to destroy a spin $s$ electron and create a spin $s$ hole cannot
be non-zero, even in the proximity of the superconductor. The
Boguliobov amplitudes $\scxone{d_{ik-s}^\dagger b_{iks}^\dagger}$
and $\scxone{b_{iks}d_{ik-s}}$ decay exponentially fast (with a
length scale set by the ferro$-$anti-ferro gap) in the
ferromagnetic wire. So, in our $S$-matrix, $r_A$ is zero and there is no pair
potential due to the proximity effect in ferromagnetic wire. Also
as mentioned earlier, we must keep in mind that the influence of
the ferromagnet on  the spectrum of the superconductor has to be
negligibly small. This will be true only if the superconductor is
large enough.
Hence, for such junctions, the renormalization of the $S$-matrix
is only due to the Friedel oscillations.
Also note that in these wires, since the bulk does not have both
the spin species, $g_1$ and $g_2$ do not get renormalized. All the
cases mentioned above will therefore involve the full $4N \times
4N$ $S$-matrix since there is no reduction in number of
independent elements of the $S$-matrix which can occur when
symmetries are imposed.

\subsection{Three$-$Wire$-$Junction $-$ A Beam Splitter}
\label{ss:wirg_bs}

In this subsection, we consider the standard beam splitter
geometry comprising of a superconductor at the junction of three
quantum wires. In this case, we show that there is a fixed point that is
analogous to the Andreev fixed point of the \ns junction. The
$S$-matrix representing this fixed point is symmetric under all
possible permutations of the three wires and allows for the
maximum Andreev transmission (in all channels simultaneously
within unitarity constraints). The $S$-matrix is given by $r_A =
-1/3$ and $t_{A}=t_{A}^\prime=2/3$ with $r=t=t'=0$. We refer to
this fixed point as the {\textsl{Andreev$-$Griffith's fixed
point}} (\agfpd)~\footnote{The Griffith's fixed point represents
the most symmetric $S$-matrix for a normal three wire junction. It
is given by $r=-1/3$ and $t=2/3$ where $r$ is the reflection with
in each wire and $t$ is the transmission from one wire to the
other. The boundary condition for the three wire junction
corresponding to the above mentioned $S$-matrix was obtained by
Griffith~\cite{griffith} hence we refer to it as the Griffith's
fixed point.}.

For an analytic treatment of this case, we will consider a
simplified situation where there is a complete symmetry between
two of the wires, say $1$ and $2$, and the $S$-matrix is
real. In addition, the elements of the $S$-matrix corresponding to
transmission or reflection of an incident electron (hole) to a
reflected or transmitted electron (hole) are set to zero so that
only Andreev processes participating in transport. Then the
$S$-matrix is given by
%
\beq S =
\begin{vmatrix} ~0 & 0 & 0 &
r_{A} & t_{A} & t_{A}^\prime~\\
~0 & 0 & 0 & t_{A} & r_{A} & t_{A}^\prime ~\\
~0 & 0 & 0 & t_{A}^\prime & t_{A}^\prime & r_{A}^\prime~ \\
~r_{A} & t_{A} & t_{A}^\prime & 0 & 0 & 0~\\
~t_{A} & r_{A} & t_{A}^\prime & 0 & 0 & 0~\\
~t_{A}^\prime & t_{A}^\prime & r_{A}^\prime & 0 & 0 & 0~\\
 \end{vmatrix}
\label{smatbs} \eeq
\noindent Here, $r_A$ and $t_A$ and $t'_A$ are real parameters
which satisfy~\cite{das}
\begin{eqnarray}
t_A&=&1+r_A~, \non\\ r'_A&=&-1-2r_A~,
\non\\
t'_A&=&\sqrt{(-2r_A)(1+r_A)}~, \non\\{\mathrm{and}} \quad  -1 &\le
& r_A \le 0 \label{unit}
\end{eqnarray}
 by unitarity. Using Eq.~\ref{unit}, the simplified \rg equation
for the single parameter $r_A$ is given by
 \begin{equation}
\dfrac{dr_A}{dl} =  \alpha'~ \left[r_A  (1+r_A) (1+3r_A) \right]
\end{equation}
%
\begin{figure}[htb]
\begin{center}
\epsfig{figure=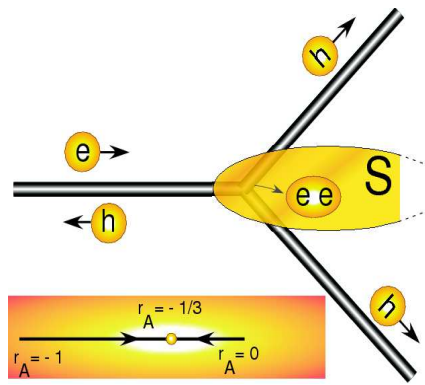,width=6cm,height=5cm}
\end{center}
\caption{{Schematic representation of the situation where a
three$-$wire$-$junction is hooked to the stable fixed point,
\agfpd. An incident electron in one wire is either reflected back
as a hole in the same wire or is transmitted as a hole in another
wire along with the addition of the two electrons into the
superconductor forming a Cooper pair. The direction of \rg flow
from two unstable fixed points to the stable fixed point (\agfpd)
is also depicted on the bottom left side of the diagram.}}
\label{nsnfig5}
\end{figure}
So, within the real parametrization we have two unstable fixed
points, given by $r_A=0$ and $r_A=-1$ and a stable fixed point
given by $r_A=-1/3$. The $r_A=0$ fixed point corresponds to a
situation where there is perfect \car between wires $1$ and $2$
and wire $3$ gets cut off from the remaining two wires (labelled
by $1$ and $2$)
and is in the perfect \ar limit with the superconductor.
The $r_A=-1$ fixed point corresponds to a
situation where all the three wires are disconnected from each
other and are in perfect \ar limit individually with the
superconductor. The third fixed point given by $r_A=-1/3$
corresponds to a perfect Andreev limit of the three wire junction
where an incident electron is either Andreev-reflected into the
same wire as a hole or is transmitted as a hole via \car into
another wire. This is essentially the \agfpd. It is very
interesting to note that the original Griffith's fixed point was a
repulsive fixed point~\cite{lal,das} whereas the \agfp is an
attractive fixed point. This can be understood as follows.
Here, there is no scattering from the Friedel oscillations as the
junction is assumed to be reflection-less, whereas there exists a
proximity induced pair potential, which induces an effective
attractive interaction between the electrons. Hence, the physics is
very similar to the well-known
Luttinger Liquid physics, which says that for attractive interaction
between the electrons, back-scattering is an irrelevant operator.
Hence the stable fixed point here will be the one which will have
maximal transmission between the wires. So, it is not surprising
that the \agfp turns out to be a stable fixed point. Thus, for a
reflection-less junction, we have found a stable fixed point with
intermediate transmission and reflection.

\section{\label{sec:five}Results}
In this section, we will consider various physical cases and see
what the \rg flows mean for the conductances in each case.

\subsection{\ns Junction} \label{ss:results_ns}

\begin{figure}[htb]
\begin{center}
\epsfig{figure=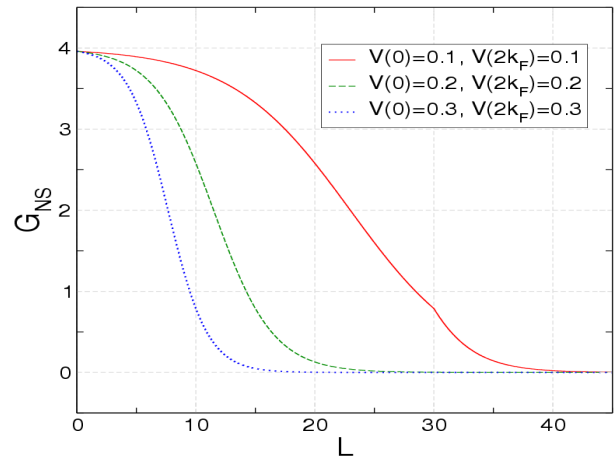,width=8cm,height=6cm}
\end{center}
\caption{Conductance of the \ns junction is plotted in units of $e^2/h$
as a function of the dimensionless parameter $l$ where
$l=ln(L/d)$ and $L$ is either $L_{T}=\hbar v_{F}/k_{B} T$ at zero bias
or $L_{V}=\hbar v_{F}/eV$ at zero temperature and $d$ is the short
distance cut-off for the \rg flow. The three curves correspond to
three different values of $V(0)$ and $V(2k_{F})$.}
\label{nsnfig6}
\end{figure}

First, we give the results for the \ns junction, just to contrast
with the results of the \nsn junction. Here, we have only two
parameters, $r$ and $r_A$. The conductance occurs only due to the
\ar amplitude, $r_A$ which obeys the \rg equation given by
Eq.~\ref{rans}. As mentioned already, there is no flow of the
particular linear combination of the interaction parameters $2g_2 -
g_1$ that occurs in the equation and the \rg flow of the conductance
is therefore monotonic. The conductance as a function of the length
scale for different interaction parameters $V(0)$ and $V(2k_{F})$ is
plotted in Fig.~\ref{nsnfig6}. $L$ here simply denotes the length at
which the \rg is cut-off. So if we take very long wires $L_W \gg
L_T$, then the cut-off is set by the temperature, and the plot shows
the variation of the conductance as a function of $L_T$ starting
from the high temperature limit, which here is the superconducting
gap $\Delta$. We observe that as we lower the temperature, the
Andreev conductance decreases monotonically. Also it was established
in Ref.\onlinecite{jap3} that the power law scaling of conductance
($~|r_{A}|^{2}$) calculated from \wirg and bosonization were  found
to be in  agreement with each other for the limiting cases of
$|r_{A}|^{2} \cong 1$ and $~|r_{A}|^{2}\cong 0$ ( which are the only
limits where bosonization results are valid) provided effects due to
electron-electron induced back-scattering in the wires is neglected.

\subsection{Ballistic \nsn Junction}
\label{ss:results_nsnb}
In this subsection, we consider the case of a reflection-less
ballistic junction between the superconductor and the two wires,
\ie~$r=0$. This implies that the renormalization of the $S$-matrix
due to the Friedel oscillations is absent. The only
renormalization is  due to reflections from the proximity effect induced
pair potential.
Let us now consider various interesting cases:\\
\begin{enumerate}
 \item[] {\textsl{(a)}} $r = 0, r_A = 0, t \neq 0, t_A \neq 0$.
In this case, since we have both $r = 0$ and $r_A = 0$, there is
no \rg flow of the transmission and the conductance is frozen at
the value that it had for the bare $S$-matrix. The most
interesting situation in this case arises when $t = t_A$. For this
case, the probability for an incident electron in one wire, to
transmit in the other wire as an electron due to $t$ or as hole
due to $t_A$ is equal, leading to perfect cancellation of charge
current.
\item[] {\textsl{(b)}} $r = 0, t = 0, r_A \neq 0, t_A \neq 0$. For
this case, one can easily check from the \rg equations
(Eqs.~\ref{rnsn}-\ref{tansn}) that if we start our \rg flow with
the given parameters at high energies, then the value of $r,t$
remain stuck to the value zero under the \rg flow. Hence, in this
case the two parameter subspace $r_A \neq 0, t_A \neq 0$ remains
secluded under the \rg flow. The \rg equation for $t_A$ is given
by
 \begin{eqnarray}
 \dfrac{dt_A}{dl} &=& \alpha' \, t_A \, (1 - |t_A|^2)
\label{33}
 \end{eqnarray}
The above equation can be integrated to obtain an expression for
\car probability ($T_A = |t_A|^2$),

\bea T_{A}(L)=\frac{T_{A}^{0} \left[[1+2\alpha_{1}
\ln(\frac{L}{d})]^\frac{3}{2}(\frac{d}{L})^{-(2\alpha_{2}
-\alpha_{1})}\right]}{R_{A}^{0}+T_{A}^{0}\left[[1+2\alpha_{1}
\ln(\frac{L}{d})]^\frac{3}{2}(\frac{d}{L})^{-(2\alpha_{2}
-\alpha_{1})}\right]} \label{ta} \eea $T_A^0$ and $R_A^0$ are the
\car and \ar probabilities at the short distance cut-off, $L=d$.
We notice that the \rg equation and its solution are very similar to
that for the single scatterer problem~\cite{matveev} apart from a
sign difference on the RHS of the equation and the dependance of
the interaction parameter $\alpha'$ on $g_1$ and $g_2$. This implies
that even if we start with a small crossed Andreev
transmission across the junction, the
\rg flow will take us towards the limit of perfect transmission. This is
in sharp contrast to the normal transmission across a single
scatterer. For the single barrier
problem, the equation for the \rg flow of $t$
was
by
 \begin{eqnarray}
 \dfrac{dt}{dl} &=&- \alpha \, t \, (1 - |t|^2).
 \end{eqnarray}
Hence, $t=0$ was the stable fixed point. But if the electron-electron
interactions had been attractive, then the sign on the RHS would have
been positive and $t=1$ would have been the stable fixed point.
Thus, the \rg flow of $t_A$  for the case when $r_A\ne 0,t=r=0$, and repulsive
interactions, is very similar to the \rg flow for $t$ when $r\ne 0,t_A=r_A=0$
but with attractive interactions.
In both
cases transmission is relevant and $t=1$ and $t_{A}=1$ are the stable fixed
points. On the other hand the \rg flow of $t_{A}$ for the case of
$r_{A} \neq 0, t=0, r=0$ and attractive electron-electron interaction
($V(0), V(2k_{F})<0$) in the wire is very similar to the \rg flow for $t$
for the case $r \neq 0, r_{A}=0, t_{A}=0$ and repulsive electron-electron
interaction ($V(0), V(2k_{F})>0$). In both cases transmission
is irrelevant and $t=0$ and $t_{A}=0$ are the stable fixed points.
At an intuitive level, one can perhaps say that even if we start
with repulsive inter-electron interactions, the proximity-induced pair
potential leads to a net attractive interaction between the electrons,
which is responsible for the counter-intuitive \rg flow.

Also notice that while solving the above \rg equation for $t_A$, we have to
take into account the \rg flow of the interaction parameter
($\alpha^\prime$) itself. This will lead to
non-power law (non Luttinger) behavior for the conductance close
to $|t_A|\,\simeq\,0$ or $|t_A|\,\simeq\,1$. It is worth pointing
out that the non-power law part appearing in Eq.~\ref{ta} is
identical to Ref.~\onlinecite{matveev}, even though the
interaction parameter for their case was proportional to
$g_2-2g_1$ and for our case it is $g_2+g_1$.
But of course this will not lead to any non-monotonic
behavior as $\alpha^\prime$ can not change sign under \rg flow. So
the stable fixed point for this case is the \cafpd.
\item[] {\textsl{(c)}} $r = 0, t_A = 0, r_A \neq 0, t \neq 0$.
This case is identical to the case (b) discussed above except for
the fact that we have to replace $t_A$ in the previous case by
$t$. In this case also the two parameter subspace $r_A \neq 0,
t\neq 0$ remains secluded under \rg flow. The \rg equation for $t$
is given by
 \begin{eqnarray}
 \dfrac{dt}{dl} &=& \alpha' t (1 - |t|^2)
 \end{eqnarray}
Here also, $t=1$ remains the stable fixed point and $t=0$ is the
unstable fixed point.
\item[] {\textsl{(d)}} $r = 0, t \neq 0, r_A \neq 0, t_A \neq 0$.
In this case if we start from a symmetric situation, \ie~$t=t_A$,
we can see from the \rg equations in Eqs.~\ref{tnsn} and
\ref{tansn} that both $t$ and $t_A$ have identical \rg flows. So,
the sub-gap conductance $G = G_{CA} - G_{CT}$ vanishes identically
and remains zero through out the \rg flow. Hence this $S$-matrix
can facilitate production of pure \scd~\cite{drsahaepl} if we
inject spin polarized electrons from one of the leads as the
charge current gets completely filtered out at the junction.
\end{enumerate}








\subsection{Ballistic \fsf Junction}
\label{ss:results_fsf}
Here, we consider the case where both the wires are spin
polarized. In this case we have two interesting possibilities,
\ie~ either both the wires have aligned spin polarization or they
have them anti-aligned. In either case the Andreev reflection
amplitude is zero on each wire due to reasons explained earlier.
\\
\begin{enumerate}
\item[]{\textsl{(a)}} When the two wires have their spins aligned,
$t \ne 0$, but $t_A=0$ because for \car to happen we need up(down)
spin polarization in one wire and down(up) spin polarization on
the other wire which is not possible in this case.
\item[]{\textsl{(b)}} When the two wires have their spins
anti-aligned, $t=0$, but $t_A\ne 0$ because the up(down) electron
from one wire can not tunnel without flipping its spin into the
other wire. As there is no mechanism for flipping the spin of  the
electron at the junction, such processes are not allowed.
\end{enumerate}
Hence these two cases can help in separating out and measuring
amplitudes of the direct tunneling process and the \car process
experimentally~\cite{beckmann}. Both these are examples of case
II, since they have both $r=0$ and $r_A=0$. In this case, neither
$t$ nor $t_A$ change under \rg flow and hence conductance is not
influenced by electron-electron interaction at all.

\subsection{Non-ballistic \nsn Junction}
\label{ss:results_nsnnb}
\subsubsection{Without \ar on individual wires}

Here we consider an \nsn junction with finite reflection in each
wire and no \ar in the individual wires. So the renormalization of
the $S$-matrix is purely due to Friedel oscillations and there are
no contributions coming from scattering due to the pair
potential. Below we discuss two cases : \\

\begin{enumerate}

\item[]{\textsl{(a)}} $r_A = 0, t = 0, r \neq 0, t_A \neq 0$.
This is an example of case II mentioned in
Subsection.~\ref{ss:wirg_nsn}.
The \rg equations (Eqs.~\ref{rnsn}-\ref{tansn}) predict that
$r_A,t$ will remain zero under the \rg  flow and $r,t_A$ form a
secluded sub-space. The \rg equation for this case is given by
\begin{eqnarray}
\dfrac{dt_A}{dl} &=& -\,\alpha\, t_A (1 - |t_A|^2)
\end{eqnarray}
Note the change in sign on the RHS with respect to the \rg
equation for $t_A$ (Eq.~\ref{33}) for the ballistic case. This
change in sign represents the fact that the ballistic case
effectively represents a situation corresponding to attractive
electron-electron interaction while  this case  corresponds to a
purely repulsive electron-electron interaction.

 Fig.~\ref{nsnfig7} shows the behavior of
conductance ($G_{CA}$) for this case. The conductance in the main
graph shows a non-monotonic behavior.

%
\begin{figure}[t]
\begin{center}
\epsfig{figure=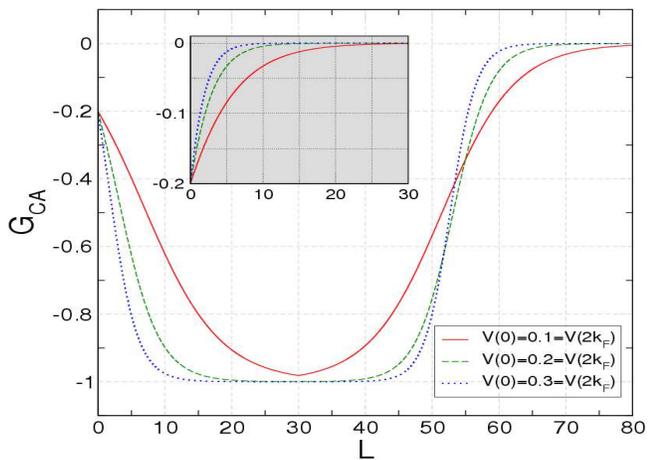,width=8.5cm,height=6cm}
\end{center}
\caption{Conductance $G_{CA}$ of the \nsn junction is plotted (when the
two leads have anti-parallel spins) in units of $e^2/h$ as a function of
the dimensionless parameter $l$ where $l=ln(L/d)$ and $L$ is either
$L_{T}=\hbar v_{F}/k_{B} T$ at zero bias or $L_{V}=\hbar v_{F}/eV$ at zero
temperature and $d$ is the short distance cut-off for the \rg flow.
The three curves correspond to three different values of $V(0)$
and $V(2k_{F})$. The inset shows the behavior of the same conductance for
fixed values of $\alpha$.}
\label{nsnfig7}
\end{figure}
%
To contrast, we also show in the inset, the behavior when the
renormalization of $\alpha$ in not taken into account. Thus, it is
apparent from the plot that the non-monotonicity is coming solely
from the \rg evolution of $\alpha$. The inset and the main graph,
both start from the same value of $t_A$. Even though this case is
theoretically interesting to explore, its experimental realization
may not be viable.
%
This is because of the following reasons. Here we have $r_A = 0$
on both wires, which can only happen if the wires are
ferromagnetic. However, we also know that if the wires are
ferromagnetic, there is no scaling of $\alpha$ parameter and hence
there will be no interesting non-monotonic trend in the conductance.
So it is hard to find a physical situation where $r_A = 0$ and at
the same time, there is renormalization of the interaction
parameter $\alpha$. Lastly note that the conductance $G_{CA}$ is
negative. The process responsible for the conductance,
(\ie~\card), converts an incoming electron to an outgoing hole or
vice-versa, resulting in the negative sign.
\item[]{\textsl{(b)}} $r_A = 0, t_A = 0, r \neq 0, t \neq 0$. This
case is identical to the previous case with the replacement of
$t_A$ by $t$. Fig.~\ref{nsnfig8} shows the the \ct conductance
$G_{CT}$ as a function of the length scale. It shows a similar
non-monotonic behavior with positive values for the conductance.
The inset shows the behavior of $G_{CT}$ when the renormalization
of $\alpha$ in not taken into account.
%
\begin{figure}[htb]
\begin{center}
\epsfig{figure=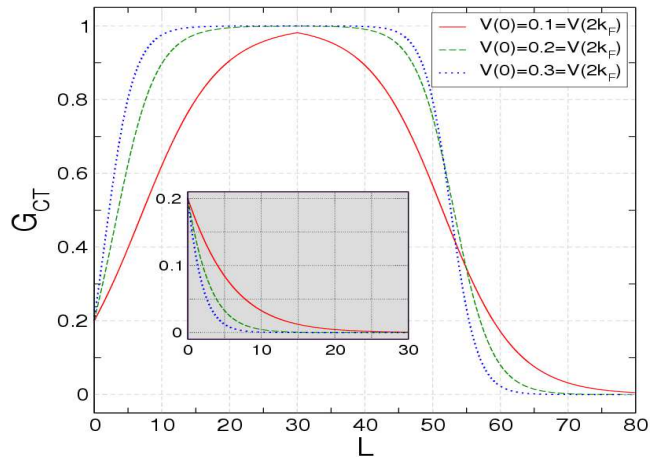,width=8.5cm,height=6cm}
\end{center}
\caption{Conductance $G_{CT}$ of the \nsn junction (when the
two leads have parallel spins) in units of $e^2/h$ as a function of
the dimensionless parameter $l$ where $l=ln(L/d)$ and $L$ is either
$L_{T}=\hbar v_{F}/k_{B} T$ at zero bias or $L_{V}=\hbar v_{F}/eV$ at zero
temperature and $d$ is the short distance cut-off for the \rg flow.
The three curves correspond to three different values of $V(0)$
and $V(2k_{F})$. The inset shows the behavior of the same conductance for
fixed values of $\alpha$.}
\label{nsnfig8}
\end{figure}

\subsubsection{With \ar on individual wires}

\item[] {\textsl{(a)}} $r_A \neq 0, t_A \neq 0, r \neq 0, t \neq
0$. This is the most interesting case, where both $r$ and $r_A$
are non-zero, and we get an interplay of the effects due to
scattering from Friedel oscillations and from
proximity induced pair potential. Here, all the four parameters
are non-zero and flow under \rgd, as do the interaction parameters
$\alpha$ and $\alpha'$. An example where the system starts in the
vicinity of the unstable fixed point {\textsf{SFP}} (as mentioned
in Case IV in the Subsection.~\ref{ss:wirg_nsn}) is shown in
Fig.~\ref{nsnfig9}. The \nsn conductance here is defined as
$G_{NSN} = G_{CA} - G_{CT}$. Here also we observe a strong
non-monotonicity in the conductance which comes about due to
interplay of the electron and the hole channels, which contribute
to the conductance with opposite signs, coupled with the effects
from the \rg flow of the interaction parameters.

\end{enumerate}

\subsection{Non-ballistic \fsn Junction}
\label{ss:results_fsnnb}

In this case, for the ferromagnetic wire $r_A = 0$, but for the
normal wire $r_A$ has a finite value. As explained earlier, the
interaction parameters $\alpha$ and $\alpha'$ on the ferromagnetic
side do not renormalize, whereas they do on the normal side.
Hence, even if we start from a situation where the interaction
parameter $\alpha$ and $\alpha'$ are symmetric for the two wires,
\rg flow will always give rise to an asymmetry in the interaction
strength. Therefore, it becomes a very interesting case to study
 theoretically. The $S$-matrix for this
case has neither spin up-spin down symmetry, nor the wire index
(left-right for two wires) symmetry. Only the particle-hole
symmetry can be retained while parameterizing the $S$-matrix. This
case gets very complicated to study theoretically because the
minimum number of independent complex-valued parameters that are
required to parameterize the $S$-matrix is nine as opposed to
four in the
\nsn case. These are given by
$r_{\up\up}^{11}, ~r_{\up\up}^{22}, ~r_{\dn\dn}^{22},
~t_{A\up\up}^{12}, ~t_{A\dn\dn}^{21}, ~r_{A\up\up}^{22},
~r_{A\dn\dn}^{22}, ~t_{\up\up}^{12}, ~{\mathrm{and}}~
t_{\up\up}^{21}$.
 Here, $1$($2$) is the wire index for the ferromagnetic
 (normal) wire while, $\up$ and $\dn$ are the respective
 spin polarization indices for the electron.

\begin{figure*}[htb]
\hskip -8cm \epsfig{figure=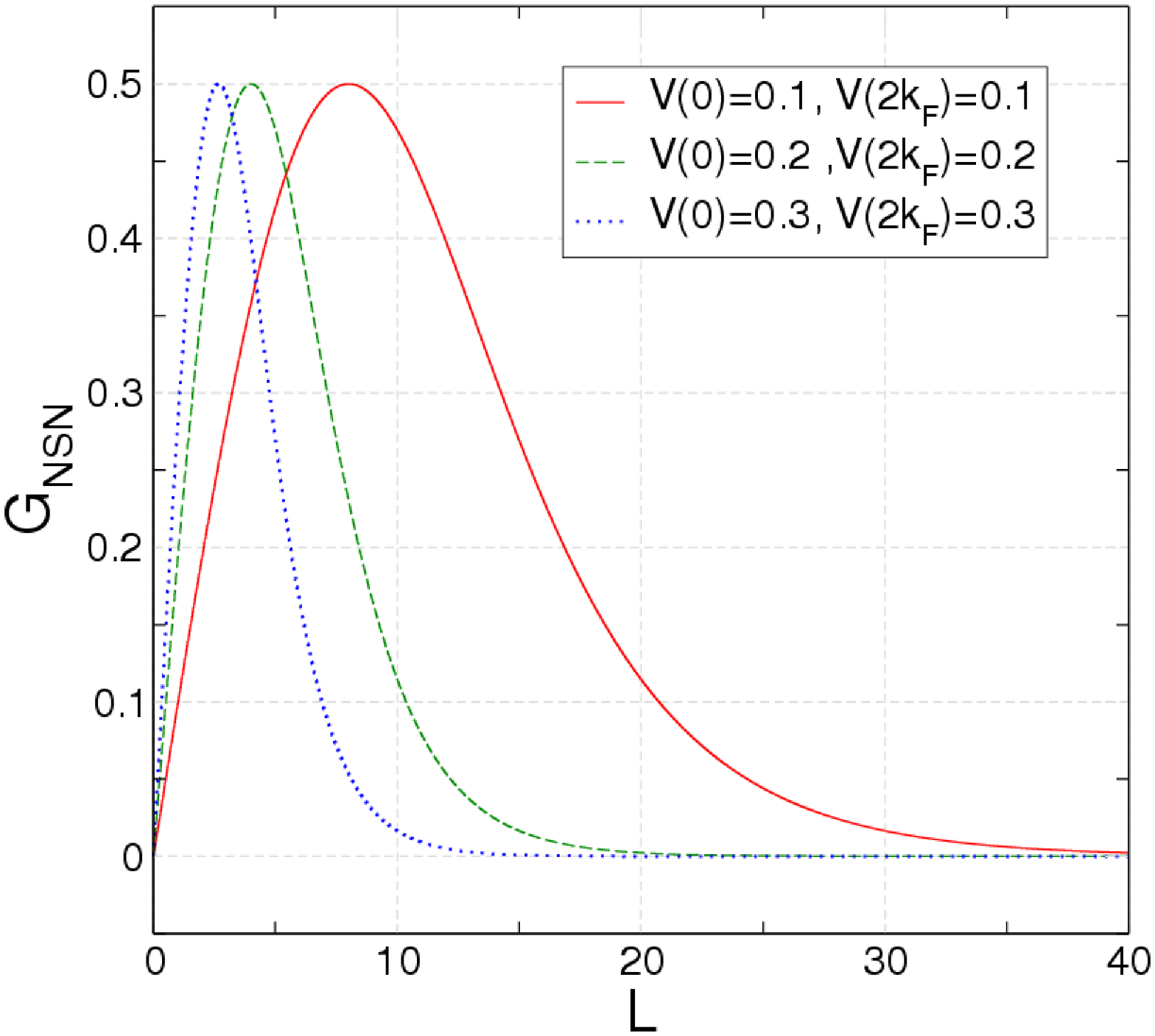,width=8cm,height=6cm}
\vskip -6cm \hskip 9cm
\epsfig{figure=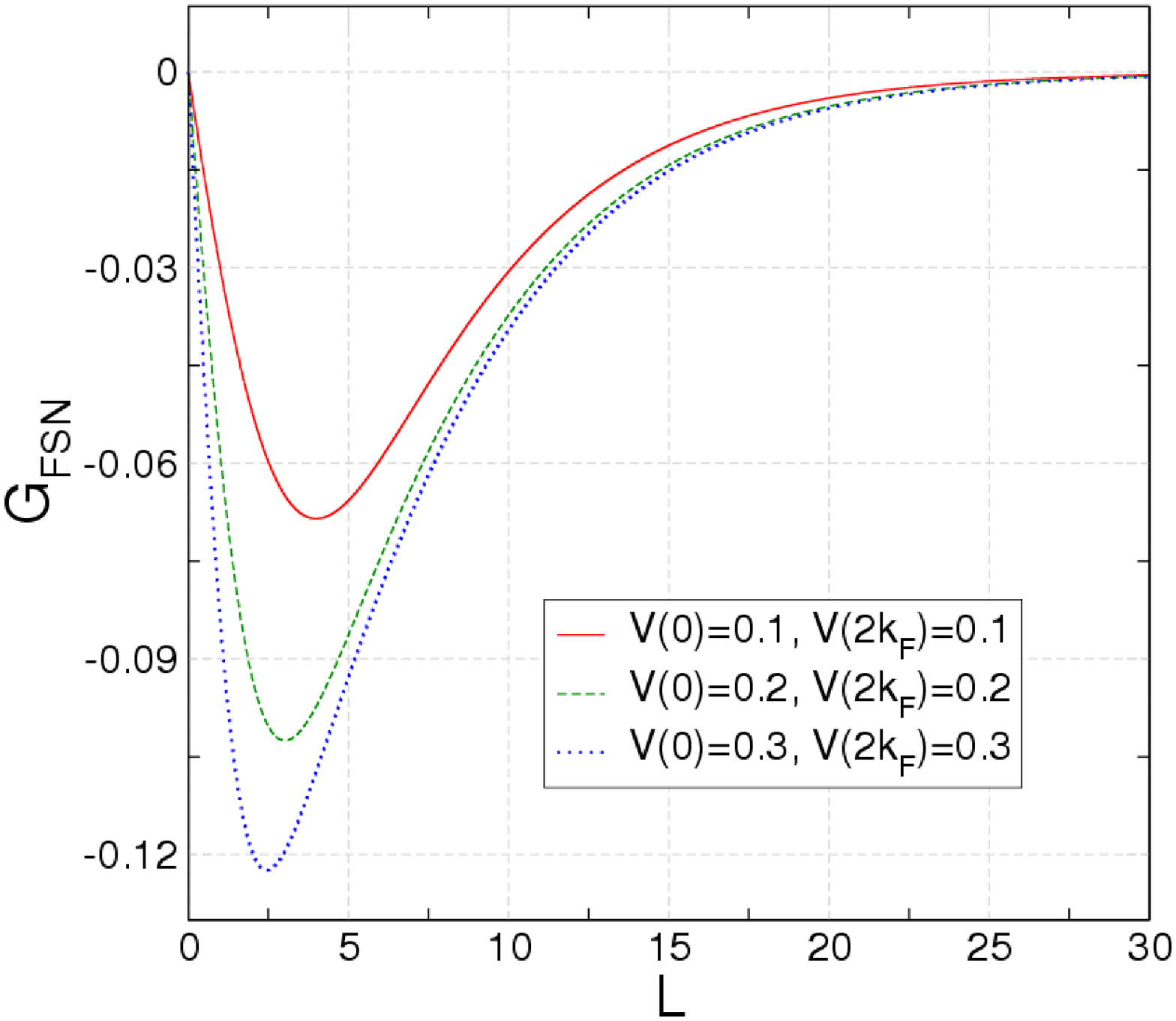,width=8cm,height=6cm}
\caption{Left: Conductance of the \nsn junction $G_{NSN}=|t_A|^2-|t|^2$ is
plotted in units of $2e^2/h$ as a function of the dimensionless parameter
$l$ where $l=ln(L/d)$ and $L$ is either $L_{T}=\hbar v_{F}/k_{B} T$ at zero
bias or $L_{V}=\hbar v_{F}/eV$ at zero temperature and $d$ is the short
distance cut-off for the \rg flow. The three curves correspond to three
different values of $V(0)$ and $V(2k_{F})$.
Right: Charge conductance $G_{FSN}=|t_A|^2-|t|^2$ is plotted for \fsn case
as a function of the dimensionless parameter $l$ where $l=ln(L/d)$ and $L$
is either $L_{T}=\hbar v_{F}/k_{B} T$ at zero bias or $L_{V}=\hbar
v_{F}/eV$ at zero temperature in units of $e^2/h$ and $d$ is the short
distance cut-off for the \rg flow. The three curves correspond to three
different values of $V(0)$ and $V(2k_{F})$.}
\label{nsnfig9}
\end{figure*}

So, the minimal $S$-matrix representing the \fsn
junction is given by
\beq S ~=~
\begin{vmatrix}
~r &   t & 0 &  0 & t_{A} & 0~ \\
~t^\prime &  r^\prime & 0 &  0 & r_{A} & 0~\\
~0 &  0 & r^{\prime \prime} & t_{A}^\prime & 0 & r_{A}^\prime~\\
~0 &  0 & t_{A} &  r & 0 & t~\\
~t_{A}^\prime &  r_{A}^\prime &  0 & 0 & r^{\prime \prime} & 0~\\
~0 &  0 & r_{A} &  t^\prime & 0 & r^{\prime}~\\
 \end{vmatrix}
\label{smat4} \eeq
%
%
%
%
The \rg equations for the nine independent parameters are given in
Appendix~\ref{appa}. We write down a representative $S$-matrix
which satisfies all the constraints of the \fsn junction and unitarity,
and study
its \rg flow numerically by solving the nine coupled differential
equations. The modulus of the $S$-matrix elements are given by  $
|r_{\up\up}^{11}|=
|r_{\up\up}^{22}|=|r_{\dn\dn}^{22}|=|t_{A\up\up}^{12}|=
|t_{A\dn\dn}^{21}|=|r_{A\up\up}^{22}|=|r_{A\dn\dn}^{22}|
=|t_{\up\up}^{12}|=\lvert t_{\up\up}^{21}\rvert=1/\sqrt{3}$
and the corresponding phases associated with each of these
amplitudes are $\pi/3,\pi,0,-\pi/3,0,\pi/3,0,\pi,-\pi/3$
 respectively. Here also we observe a non-monotonic behavior of
 conductance, $G_{FSN}$ as a function of $l$ as shown in
 Fig.~\ref{nsnfig9}.

%

\subsection{Non-ballistic \fsf junction}
\label{ss:results_fsfnb}

 Here, we will consider the case where both the wires are spin polarized.
 This case is similar to the ballistic
 case. Since here $r \ne 0$ and $r_A = 0$, we will have the usual Friedel
 oscillations and the conductance will go to zero as a power law.
 Here again we have two instructive cases:
%
\begin{enumerate}
\item[]{\textsl{(a)}} When the two wires connected to the
superconductor have their spin polarization aligned, \ie~$t \ne
0$, but $t_A=0$ and
\item[]{\textsl{(b)}} When the two wires have their spin
polarization anti-aligned, $i.e.$, $t=0$ but $t_A\ne 0$.
\end{enumerate}
Both these are examples of case II of
Subsection.~\ref{ss:wirg_nsn}. Either $t$ or $t_A$ need
to be zero in the two cases mentioned above. Hence the parameters
which are zero will remain zero under \rg, while the non-zero
parameters will flow according to Eqs.~\ref{tnsn} and \ref{tansn}
respectively. The conductances are the same as in the \nsn case
except that the interaction parameters cannot flow now.
This has already been plotted in the insets in Figs.~\ref{nsnfig7}
and \ref{nsnfig8}.
Since the electrons are now effectively spin-less, $\alpha$ and
$\alpha'$ do not flow, and we get a monotonic fall-off of the
conductance in both the cases.\\
The results of this section are summarized in the table below.\\

\vskip -0.5cm
\begin{widetext}
\hskip +0.7cm
\begin{tabular}{|c|c|c|c|c|c|c|}
\hline
$t$ & $t_{A}$ & $r_{A}$ & $r$ & Stability & Intermediate fixed point &
Relevant physics\\ \hline
0 & 0 & 0 & 1 & Stable & $\times$ &  \rfpd \\ \hline
0 & 0 & 1 & 0 & Unstable & $\times$ &  \afpd \\ \hline
0 & 1 & 0 & 0 & Unstable & $\times$ &  \cafpd \\ \hline
1 & 0 & 0 & 0 & Unstable & $\times$ &  \tfpd \\ \hline
1/2 & 1/2 & -1/2 & 1/2 & Unstable & $\surd$ & \sfpd, Non-monotonic charge
current\\ \hline
$e^{i \phi_{1}}\sin \theta$ & $e^{i \phi_{2}}\cos \theta$ & 0 & 0 &
Marginal & $ - $ & Pure spin current when $t=t_{A}$ \\ \hline
\end{tabular}
\vskip +0.2cm
\end{widetext}
\vspace{0.2cm}

\section{\label{sec:six} Generalization to the case of three wires}

\begin{figure}[htb]
\begin{center}
\epsfig{figure=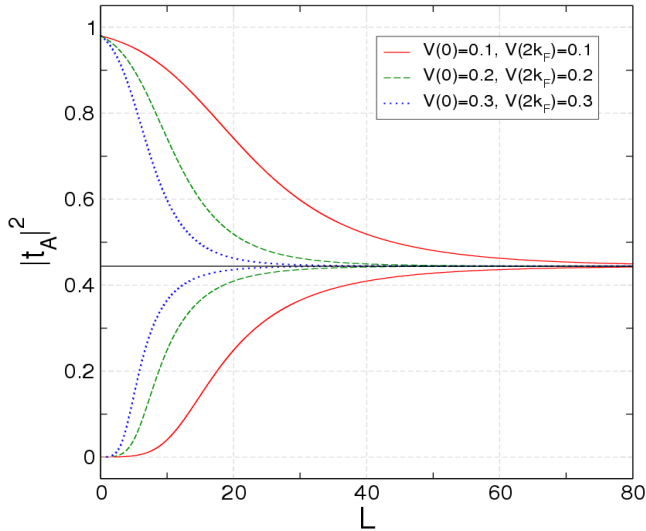,width=8.5cm,height=7cm}
\end{center}
\caption{$|t_A|^2$ is plotted as a function of dimensionless parameter
$l$ where $l=ln(L/d)$ and $L$ is either $L_{T}=\hbar v_{F}/k_{B} T$ at zero
bias or $L_{V}=\hbar v_{F}/eV$ at zero temperature and $d$ is the short
distance cut-off for the \rg flow. The three curves correspond to three
different values of $V(0)$ and $V(2k_{F})$.
The set of curves in the top represent the \rg flow of $|t_A|^2$ when the
starting point is in the vicinity of $r_A = 0$ fixed point while, the set
of curves in the bottom half represent the \rg flow of $|t_A|^2$ with
starting point close to $r_A = -1$.}
\label{nsnfig10}
\end{figure}

In this section, we consider the case of three wires connected to
a superconductor.
We assume that all the wires are connected within the phase
coherence length of the superconductor. Hence, \car can occur by
pairing the incident electron with an electron from any of the
other wires and emitting a hole in that wire. The conductance
matrix can hence be extended for three wires as
\beq
 \begin{vmatrix}
~I_1~ \\
~I_2~ \\
~I_3~ \\
  \end{vmatrix} = \begin{vmatrix}
~G_{r\,11} & G_{t\,12} & G_{t\,13}~  \\
 ~G_{t\,12}  & G_{r\,22} &  G_{t\,23}~ \\
 ~G_{t\,13} &  G_{t\,23} &
G_{r\,33}~
\\
\end{vmatrix}
 \begin{vmatrix}
~V_1~ \\
~V_2~ \\
~V_3~\\
   \end{vmatrix}
   \eeq with
   %
$  G_{r\,ij} = G_{{A\, ii}} + \sum_{j}(G_{{CA\,ij}}+G_{{CT\,ij}})$
and
$ G_{t\,ij} = G_{{CA\,ij}} - G_{{CT\,ij}}$ and the generalization
to $N$ wires is obvious. Note that the conductances $G_{CA\,ij} =
G_{CA\,ji}$ and $G_{CT\,ij} = G_{CT\,ji}$. The relations of the
conductances to the reflections and transmissions is obvious, for
e.g., $G_{A\,ii} \propto |r_{A\,ii}|^2$ as before while
$G_{CT\,ij} \propto |t_{ij}|^2$ and $G_{CA\,ij} \propto
|t_{A\,ij}|^2$. The \rg equations for the three wire case can be
written using the matrix equation as given in Eq.~\ref{smat1}
except that the $S$ matrix is now $12 \times 12$ dimensional. For
a system with particle-hole, spin up-spin down and wire index
symmetry, the $S$ matrix is given by, \beq S_\uparrow =
S_\downarrow =
\begin{vmatrix}
~r & t & t^\prime & r_{A} & t_{A} & t_{A}^\prime~\\
~t &r&t & t_{A} & r_{A} & t_{A}~\\
~t^\prime & t & r & t_{A}^\prime &t_{A}&r_{A} ~\\
~ r_{A} & t_{A} & t_{A}^\prime &r & t & t^\prime~\\
~t_{A} & r_{A} & t_{A}& t & r& t~\\
~t_{A}^\prime & t_{A}& r_{A} &t^\prime &t& r~\\
 \end{vmatrix}
\label{smat3} \eeq
where we have chosen six independent parameters, with $t_{12} =
t_{21} = t_{23} = t_{32} = t$ and $t_{13} = t_{31} = t'$
and similarly for the \car parameter $t_A$.
%
 The $F$ matrix now generalizes to
\beq
F = \begin{vmatrix}
~\frac{\alpha r}{2}& 0 & 0 & \frac{-\alpha^\prime r_A} {2} &0 &0 ~\\
~0& \frac{\alpha r}{2}& 0 & 0 & \frac{-\alpha^\prime r_A}{2}&0  ~\\
~0& 0&\frac{\alpha r}{2}&0 & 0 & \frac{-\alpha^\prime r_A}{2} ~\\
~\frac{-\alpha^\prime r_A}{2} & 0 & 0& \frac{\alpha r}{2}& 0&0 ~\\
~0& \frac{-\alpha^\prime r_A}{2} & 0 & 0 & \frac{\alpha r}{2}&0 ~\\
~0&0&\frac{-\alpha^\prime r_A}{2} & 0 & 0 & \frac{\alpha r}{2} ~\\
\\
\end{vmatrix}
\label{fmat2} \eeq and the \rg equations for the six independent
parameters are given in Appendix~\ref{appb}. There exists
possibility of many more non-trivial fixed points in this case.
For instance, the \agfpd, as mentioned in
Subsection.~\ref{ss:wirg_bs}. As discussed in
Subsection.~\ref{ss:wirg_bs}, for the reflection-less case with
symmetry between just two wires, this complicated $S$-matrix takes
a very simple form, which can be dealt with analytically. Within
the sub-space considered we found that the \agfp was  a stable
fixed point. In Fig.~\ref{nsnfig10}, we show the \rg flow of
$|t_A|^2$ from two different unstable fixed points to the stable
\agfpd.

The possibility of experimental detection of such a non-trivial
fixed point with intermediate transmission and reflection is quite
interesting. From this point of view, the \agfp is a very
well-suited candidate as opposed to its counterpart, the
Griffith's fixed point~\cite{lal,das}. For a normal junction of
three \od \qwd, the $S$-matrix corresponding to $r=-1/3, t=2/3$ is
a fixed point (Griffith's fixed point), where $r$ and $t$ are the
reflection and the transmission for a completely symmetric three
wire junction.
Even though it is an interesting fixed point, it turns  out to
be a repulsive one and hence the
 possibility of its experimental detection  is very low.
On the contrary, the \agfpd, being an attractive fixed point, has
a  better possibility of  be experimentally measured. The main
point here is that even if we begin with an asymmetric junction,
which is natural in a realistic experimental situation,
the effect of interaction correlations are such that as we go down
in temperature, the system will flow towards the symmetric
junction. This can be inferred  from the results shown
in
 Fig.~\ref{nsnfig10}.

%
\section{\label{sec:seven}Summary and Discussions}
\label{sum}

To summarize, in this article we have studied transport through a
superconducting junction of multiple \od interacting quantum wires
in the spirit of the Landauer$-$Buttiker formulation. Using the \wirg
approach we derived the \rg equations for the effective $S$-matrix
and obtained the various fixed point $S$-matrices representing the
junction. In contrast to earlier \rg studies, here, we had to
include both particle and hole channels due to the proximity
effect of the superconductor. Our study led to the finding of a  novel
fixed point with intermediate Andreev transmission and
reflection even in the case of \nsn junction (\sfpd).
However,
it turns out to be unstable fixed point and hence experimentally
inaccessible. We found that transport across the superconducting
junction depends on two independent interaction parameters,
($\alpha = (g_2\,-\, 2g_1)/2\pi\hbar v_F$) which is due to the
usual correlations coming from Friedel oscillations for spin-full
electrons and ($\alpha'= (g_2 \,+\, g_1)/2\pi\hbar v_F$), which
arises due to the scattering of electron into hole by the proximity
induced pair potential in the \qwd. We computed the length scale (or
temperature) dependance of the conductance taking into account
interaction induced forward and back-scattering processes. We
found a non-monotonic dependence of the conductance for the
two-wire \nsn superconducting junction in contract to the \ns
junction where the dependence is purely monotonic.

When more than two wires are attached to the superconductor, we
found even more exotic fixed points like the \agfp which happens to
be a stable fixed point for a reflection-less symmetric junction thus
increasing its chances of being experimentally seen. But
reflection is a relevant perturbation and asymmetry between different wires
is likely to be a relevant parameter \cite{drsaha}. Hence,
the \rg flow due to these perturbations
will take the system to the \rfpd  in the low energy limit.
So, the experimental
detection of \agfp fixed point will critically depend on how efficiently
the conditions of reflectionlessness and symmetry can be maintained
in the experimental setup.
If the reflection and asymmetry are reduced to a large extent,
then as we cool the system, the S-matrix at the junction is expected to flow
very close to the \agfp fixed point. But ultimately,  the \rg flow will
lead to enhancement of any initial small value of reflection and
asymmetry and we will finally
flow to the disconnected fixed point in the $T=0$ limit. Thus  it would be an
interesting experimental challenge to look for signature of the \agfp
at intermediate temperatures.

Before we conclude, it is worth mentioning that the geometry studied in our
paper is of direct interest for the production of non-local entangled
electron pairs  propagating in two different wires.
These electron pairs are  produced by Cooper pair breaking via crossed Andreev processes when the superconductor
is biased with respect to the wires comprising the junction.
One can ask if electron-electron interaction in the wires
actually leads to enhancement of entangled electron pair production via the
crossed Andreev processes. For example, we have observed that for the
\nsn junction with $r=0, t=0, r_{A} \neq 0, t_{A} \neq0 $, interaction
can lead to enhancement of the crossed Andreev amplitude ($t_{A}$) under
\rg flow. This implies that for the case when the superconductor is
biased with respect to the wires, inter-electron interactions
enhance the production of non-local entangled pairs, for which
the amplitude $t_A$ is relevant, as compared to local entangled pairs
for which the amplitude $r_A$ would be relevant.
This is consistent with the results of Recher and Loss\cite{recher1}
who also argued that it is energetically more favourable for the
two entangled electrons of the Cooper pair to go into different wires,
rather than the same wire.
Finally the \rg flow leads to a fixed point with $t_{A}=1$ where the system
becomes a perfect entangler. A more general case would be
when $r \neq 0, t \neq 0, r_{A} \neq 0, t_{A} \neq 0$. To study this
case, one can start from the two wires \sfpd S-matrix and study the \rg
flow of $t_{A}$ for an S-matrix which is in the close vicinity of this fixed
point. The result of this study is shown in Fig.~\ref{nsnfig9}. We show
that starting from the short-distance cut-off $d$ the \rg flow initially
leads to enhancement of $t_{A}$ which will lead to an enhancement in
the production of non-local entangled pairs. Hence, these studies establish the
fact electron-electron interactions the wires can actually lead to an
enhancement of non-local entangled electron pair production.

\acknowledgements{We acknowledge the use of the Beowulf cluster at
the Harish-Chandra Research Institute in our computations. The
work of S.D. was supported by the Feinberg fellowship programme at
WIS, Israel.}

\appendix

\section{}
\label{appa}


\setcounter{equation}{0} Here we give the \rg equations for nine
independent parameters in case of a \fsn junction.
\begin{widetext}
\bea
\dfrac{dr}{dl} &=& \Big[\dfrac{\beta}{2} ~r
\left(1~-~|r|^2\right)-\dfrac{\alpha}{2}~\left(t r^{\prime \star}
t^\prime ~+~ r^{\prime\prime\star} t_A t_{A}^\prime \right)
~+~ \dfrac{\alpha \prime}{2}~ \left(tr_{A}^{\star}t_{A}^\prime ~+~
t_{A} r_{A}^{\prime\star} t^\prime \right)\Big] \label{a1}
\\
\dfrac{dt}{dl} &=& -\, \Big[ \dfrac{\beta}{2}~ |r|^2 t ~+~
\dfrac{\alpha}{2}~ \left(|r'|^{2}t~+~t_{A}r''^{
\star}r_{A}'\right)
 ~-~
\dfrac{\alpha'}{2}~\left(r_{A}^{\star}r_{A}'t~+~r'r_{A}'^{\star}t_{A}\right)\Big]
\label{a2}\\
\dfrac{dt_{A}}{dl} &=& -\,\Big[\dfrac{\beta}{2} |r|^2 t_{A}
~+~\dfrac{\alpha}{2}~ \left(|r''|^{2} t_{A} ~+~ t r^{\prime \star}
r_{A}\right)
~-~ \dfrac{\alpha'}{2}~ \left( r_{A}^{\star} r'' t ~+~ r_{A}
r_{A}^{\prime \star} t_{A} \right) \Big]
\label{a3}\\
\dfrac{dr'}{dl} &=& -\,\Big[\dfrac{\beta}{2}~r^{\star}tt' + \dfrac
{\alpha}{2}~ [r''^{\star}r_{A}r_{A}'-r'(1-|r'|^2)]
~-~ \dfrac
{\alpha'}{2}~r'(r_{A}r_{A}'^{\star}+r_{A}^{\star}r_{A}')\Big]
\label{a4}\\
\dfrac{dr_{A}}{dl} &=& -\,\Big[\dfrac{\beta}{2}~r^{\star}tt_{A} +
\dfrac{\alpha}{2}~r_{A}(|r'|^{2}~+~|r''|^{2})
~+~
\dfrac{\alpha'}{2}~(r_{A}-r_{A}^2r_{A}'^{\star}-r_{A}^{\star}r'r'')\Big]
\label{a5}\\
\dfrac{dt'}{dl} &=& -\,\Big[\dfrac{\beta}{2}~|r|^2t' + \frac
{\alpha}{2}~ (|r'|^2t'+r_{A}r''^{\star}t_{A}')
 ~-~\dfrac
{\alpha'}{2}~(r_{A}r_{A}'^{\star}t'~+~r'r_{A}^{\star}t_{A}')\Big]
\label{a6}\\
\dfrac{dr''}{dl} &=& -\,\Big[\dfrac{\beta}{2}~r^{\star}t_{A}t_{A}'
~+~ \dfrac{\alpha}{2}~ [r'^{\star}r_{A}r_{A}'-r''(1-|r''|^2)]
~-~ \dfrac
{\alpha'}{2}~r''(r_{A}r_{A}'^{\star}~+~r_{A}^{\star}r_{A}')\Big]
\label{a7}\\
\dfrac{dr_{A}'}{dl} &=&
-\,\Big[\dfrac{\beta}{2}~r^{\star}t_{A}'t~+~
\dfrac{\alpha}{2}~r_{A}'\left(|r''|^2~+~|r'|^2\right)
~+~ \dfrac
{\alpha'}{2}~\left(r_{A}'-r_{A}^{\star}r_{A}'^{2}~-~r_{A}'^{\star}r'r''\right)\Big]
\label{a8}
\\
\dfrac{dt_{A}'}{dl} &=& -\,\Big[\dfrac {\beta}{2}~|r|^2t_{A}' ~+~
\dfrac {\alpha}{2}~\left(|r''|^2t_{A}'~+~r'^{\star}t'r_{A}'\right)
~-~ \dfrac
{\alpha'}{2}\left(r_{A}'^{\star}r''t'~+~r_{A}'r_{A}^{\star}t_{A}'\right)\Big]\label{a9}
\eea
\end{widetext}

\section{}
\label{appb}


\setcounter{equation}{0} Here we give the \rg equations for six
independent parameters in case of a symmetric 3 wire \nsn
junction.
\begin{widetext}
\begin{eqnarray}%
\dfrac{dr}{dl} &=& -\,\Big[\dfrac {\alpha}{2} [ r^{\star} \left(
r_{A}^{2}~+~t^{2}+t'^{2}~+~t_{A}^{2}~+~t_{A}'^{2}\right)
~-~ r\left(1~-~|r|^{2}\right)] ~-~ \alpha'[
r|r_{A}|^{2}~+~r_{A}^{\star}\left(tt_{A}~+~t't_{A}'\right)]\Big]
\label{b1}
\\
\frac{dt}{dl} &=& -\,\Big[\alpha
\left[|r|^{2}t~+~r^{\star}\left(r_{A}t_{A}+t'^{2}~+~t_{A}'^{2}\right)\right]
~-~\alpha' \left[|r_{A}|^{ 2}
t~+~r_{A}^{\star}\left(rt_{A}~+~t't_{A}'\right)\right]\Big]
\label{b2}
\\
\dfrac{dt'}{dl} &=& -\,\Big[\dfrac {\alpha}{2}\left[2|r|^{2}t'~+~
r^{\star}\left(tt'+t_{A}t_{A}'~+~2r_{A}t_{A}'\right)\right]
~-~ \dfrac {\alpha'}{2}\{ 2|r_{A}|^{2}t'~+~r_{A}^{\star}[r
\left(t'~+~t_{A}'\right)
 ~+~\left(t't_{A}~+~tt_{A}'\right) ] \} \Big] \label{b3}
\\
 \frac{dr_{A}}{dl} &=&-\,\Big[\frac
{\alpha}{2}\{
2|r|^{2}r_{A}~+~r^{\star}\left[2tt_{A}~+~t_{A}'(t+t')\right] \}
~+~\dfrac
{\alpha'}{2}[r_{A}-r_{A}^{\star}(r^{2}+r_{A}^{2}~+~t_{A}^{2}
~+~ t_{A}'^{2}~+~2tt')]\Big] \label{b4}
\\
\dfrac{dt_{A}}{dl}&=&- \,\Big[\dfrac
{\alpha}{2}\left[2(|r|^{2}t_{A}~+~r^{\star}r_{A}t)~+~r^{\star}t't_{A}'+r_{A}^
{\star}t'^{2}\right]
~-~\dfrac {\alpha'}{2}\left[2(|r_{A}|^{2}t_{A}~+~r t r_{A}^
{\star})~+~r_{A}^
{\star}\left(t'^{2}~+~t_{A}'^{2}\right)\right]\Big] \label{b5}
\\
\dfrac{dt_{A}'}{dl} &=& -\,\Big[\frac
{\alpha}{2}\left[2\left(|r|^{2}t_{A}'~+~
r^{\star}r_{A}t')~+~r^{\star}(t't_{A}~+~tt_{A}'\right)\right]
~-~\dfrac{\alpha'}{2} [ 2\left(|r_{A}|^{2}t_{A}'~+~ r
t'r_{A}^{\star} \right)
 +r_{A}^{\star} \left(tt'~+~t_{A}t_{A}'\right)]\Big]
\label{b6}
\end{eqnarray}
\end{widetext}


\bibliographystyle{apsrev}

\bibliography{myreferences}

\end{document}

%% file: prbmacro.tex
\newcommand\beq{\begin{equation}}
\newcommand\eeq{\end{equation}}

\newcommand\bea{\begin{eqnarray}}
\newcommand\eea{\end{eqnarray}}

\newcommand\bi{\begin{itemize}}
\newcommand\ei{\end{itemize}}

\newcommand\non{\nonumber}

\newcommand\up{\uparrow}
\newcommand\dn{\downarrow}

\newcommand\hmi{{\mathcal H}_{\textsf{int}}}

\newcommand\psiou{\Psi_{O\,\up}^{}}
\newcommand\psiiu{\Psi_{I\,\up}^{}}

\newcommand\psioudg{\Psi_{O\,\up}^\dagger}
\newcommand\psiiudg{\Psi_{I\,\up}^\dagger}

\newcommand\psiod{\Psi_{O\,\dn}^{}}
\newcommand\psiid{\Psi_{I\,\dn}^{}}

\newcommand\psioddg{\Psi_{O\,\dn}^\dagger}
\newcommand\psiiddg{\Psi_{I\,\dn}^\dagger}

\newcommand\epp{e^{i\,(k\,+\,k_F)\,x}}
\newcommand\epm{e^{i\,(-k\,+\,k_F)\,x}}

\newcommand\emp{e^{-i\,(k\,+\,k_F)\,x}}
\newcommand\emm{e^{-i\,(-k\,+\,k_F)\,x}}

\newcommand\ie{{\textsl{i.e.}}}

\newcommand\odd{1{\textendash}D}

\newcommand\od{1{\textendash}D~}

\newcommand\afpd{{\textsf{AFP}}}

\newcommand\cafpd{{\textsf{CAFP}}}

\newcommand\rfpd{{\textsf{RFP}}}

\newcommand\tfpd{{\textsf{TFP}}}

\newcommand\sfpd{{\textsf{SFP}}}

\newcommand\agfpd{{\textsf{AGFP}}}

\newcommand\agfp{{\textsf{AGFP~}}}

\newcommand\ctd{{\textsf{CT}}}

\newcommand\ct{{\textsf{CT~}}}

\newcommand\wbsd{{\textsf{WBS}}}
\newcommand\sbsd{{\textsf{SBS}}}

\newcommand\wbs{{\textsf{WBS~}}}
\newcommand\sbs{{\textsf{SBS~}}}

\newcommand\wirgd{{\textsf{WIRG}}}
\newcommand\nsd{{\textsf{NS}}}
\newcommand\rgd{{\textsf{RG}}}
\newcommand\qwd{{\textsf{QW}}}

\newcommand\nsnd{{\textsf{NSN}}}
\newcommand\fsnd{{\textsf{FSN}}}
\newcommand\fsfd{{\textsf{FSF}}}

\newcommand\fsd{{\textsf{FS}}}
\newcommand\hfd{{\textsf{HF}}}

\newcommand\scd{{\textsf{SC}}}

\newcommand\card{{\textsf{CAR}}}
\newcommand\ard{{\textsf{AR}}}

\newcommand\wirg{{\textsf{WIRG~}}}
\newcommand\ns{{\textsf{NS~}}}
\newcommand\rg{{\textsf{RG~}}}
\newcommand\qw{{\textsf{QW~}}}

\newcommand\nsn{{\textsf{NSN~}}}
\newcommand\fsn{{\textsf{FSN~}}}
\newcommand\fsf{{\textsf{FSF~}}}

\newcommand\hf{{\textsf{HF~}}}

\newcommand\car{{\textsf{CAR~}}}
\newcommand\ar{{\textsf{AR~}}}

\def\ket#1{| ~#1~ \rangle}

\def\scxone#1{\langle ~#1~ \rangle}

\def\me#1#2#3{\langle ~#2\, |~#1~|\, #3~\rangle}

\newcommand\etal{{\hbox{{\textit\ et al.\/}\textit\ }}}

\def\And{{\rm and\ }}

\def\dfrac#1#2{{\displaystyle\frac{#1}{#2}}}

\newif\ifboo \boofalse

